\documentclass[aps,pra,twocolumn,superscriptaddress]{revtex4-2}
\newcommand{\diff}{\mathop{}\!\mathrm{d}}
\usepackage{multirow,amssymb,amsbsy,amsmath}
\usepackage{graphicx} 
\usepackage{subfigure}
\usepackage[colorlinks,breaklinks]{hyperref}
\makeatletter
\newcommand\org@hypertarget{}
\let\org@hypertarget\hypertarget
\renewcommand\hypertarget[2]{%
\Hy@raisedlink{\org@hypertarget{#1}{}}#2% 
  }

\usepackage{xr}  
\usepackage{braket}
\makeatother 
\hypersetup{
colorlinks=true, % 启用颜色链接
linkcolor=blue, % 内部链接颜色
filecolor=blue, % 文件链接颜色
urlcolor=blue, % URL 链接颜色
citecolor=cyan % 引用链接颜色
}

\begin{document}
 
\title{Global-scale quantum networking using hybrid-channel quantum repeaters with relays based on a chain of balloons
} 
 
\author{Pei-Xi Liu}
\author{Yu-Ping Lin}
\affiliation{Laboratory of Quantum Information, University of Science and Technology of China, Hefei 230026, China}
\affiliation{Anhui Province Key Laboratory of Quantum Network, University of Science and Technology of China, Hefei 230026, China}
\affiliation{CAS Center For Excellence in Quantum Information and Quantum Physics, University of Science and Technology of China, Hefei 230026, China}
\author{Zong-Quan Zhou}
\email{zq\_zhou@ustc.edu.cn}
\author{Chuan-Feng Li}
\author{Guang-Can Guo}
\affiliation{Laboratory of Quantum Information, University of Science and Technology of China, Hefei 230026, China}
\affiliation{Anhui Province Key Laboratory of Quantum Network, University of Science and Technology of China, Hefei 230026, China}
\affiliation{CAS Center For Excellence in Quantum Information and Quantum Physics, University of Science and Technology of China, Hefei 230026, China}
\affiliation{Hefei National Laboratory, University of Science and Technology of China, Hefei 230088, China}
\date{\today}

\begin{abstract}
Global-scale entanglement distribution has been a formidable challenge due to the unavoidable losses in communication channels. Here, we propose a novel backbone channel for quantum network based on balloon-based aerial relays. We demonstrate for the first time that the atmospheric disturbances in balloon-based channels can be almost eliminated through optimizing beam waist positions and employing a series of adaptive optics systems, which boosts the channel efficiency to -21 dB over a 10,000 km distance, outperforming satellite-based relays by 12 dB with same device parameters. We then propose a global-scale quantum networking scheme based on hybrid-channel quantum repeaters that combine ground-based quantum repeaters and balloon-based aerial relays. Servers are interconnected globally via a chain of balloons, while multiple clients link to local servers through fiber connections, facilitating rapid client switching and network scalability. Our simulations, employing state-of-the-art Eu$^{3+}$:Y$_2$SiO$_5$ quantum memories and mature entanglement sources based on spontaneous parametric down-conversion, demonstrate an entanglement distribution rate in the sub-Hertz range between clients separated by 10,000 km. This approach offers a practical path toward global quantum networking in the near future.
\end{abstract}
\maketitle

\section{INTRODUCTION}
Long-distance distributed entanglement is a fundamental resource for quantum networks, enabling applications such as entanglement-based quantum cryptography ~\cite{PhysRevLett.67.661,gisin_quantum_2007}, network-based quantum computing~\cite{doi:10.1126/science.1214707,liu_nonlocal_2024,main_distributed_2025} and distributed quantum sensing~\cite{PhysRevLett.109.070503,nichol_elementary_2022}. However, the direct distribution of entanglement through fiber channels is limited to distances of a few hundred kilometers~\cite{PhysRevLett.134.230801}. While a single low-earth orbit (LEO) satellite can extend this to approximately 1200 km~\cite{doi:10.1126/science.aan3211}, it still can’t surpass 2000 km and has limited operational windows.

%,RevModPhys.95.045006,guha_rate-loss_2015,krovi_practical_2016,lago-rivera_telecom-heralded_2021
Quantum repeaters, which divide the long-distance entanglement distribution into shorter elementary links synchronized via quantum memories, offer a viable path forward~\cite{collins_multiplexed_2007,simon_quantum_2007,RevModPhys.83.33}. However, terrestrial fiber-based implementations face severe channel attenuation, necessitating a large number of repeater nodes and stringent requirements on quantum memories that yet to be fully met by any single atomic system \cite{liu_heralded_2021,PhysRevLett.130.213601,van_leent_entangling_2022,liu_creation_2024,knaut_entanglement_2024,doi:10.1126/sciadv.adp6442}. On the other hand, proposals on satellite-based quantum repeaters~\cite{boone_entanglement_2015,gundogan_proposal_2021,gundogan_time-delayed_2024,meister_simulation_2025,chen_zero-added-loss_2023,wallnofer_simulating_2022} typically rely on immature technologies like quantum non-demolition measurements or space-based quantum memories, and are confined to point-to-point architectures, lacking the capability for quantum networking with multiple clients.

Here, we introduce a backbone quantum channel based on a chain of stratospheric balloons, achieving a channel efficiency of –21 dB over 10,000 km. Building on this aerial link, we further propose a hybrid-channel quantum networking architecture that integrates fiber-based quantum repeaters, which  unlocks continuous, high-speed entanglement distribution among multiple globally distributed users.

Conventional wisdom holds that free-space channels near Earth's surface suffer strong atmospheric disturbances. Consequently, relays based on a chain of LEO satellites~\cite{goswami_satellite-relayed_2023} and vacuum beam guide~\cite{huang_vacuum_2024} have been considered for global-scale low-loss quantum channels, but both entail exorbitant costs for construction and maintenance. In contrast, we show that ascending to the stratosphere can reduce the residual turbulence to a manageable level: the residual atmospheric disturbances can be effectively mitigated by countermeasures of optimizing the beam waist positions and utilizing a series of adaptive optics (AO) systems~\cite{cao_long-distance_2020,gruneisen_adaptive-optics-enabled_2021}. Surprisingly, we find that balloon-based relays achieve more than an order of magnitude higher channel efficiency than satellite-based relays, thanks to their significantly shorter vertical links to the ground, thus providing an efficient and realistic backbone for a global quantum network.

\begin{figure*}[t]
\centering 
\includegraphics[width=1 \linewidth]{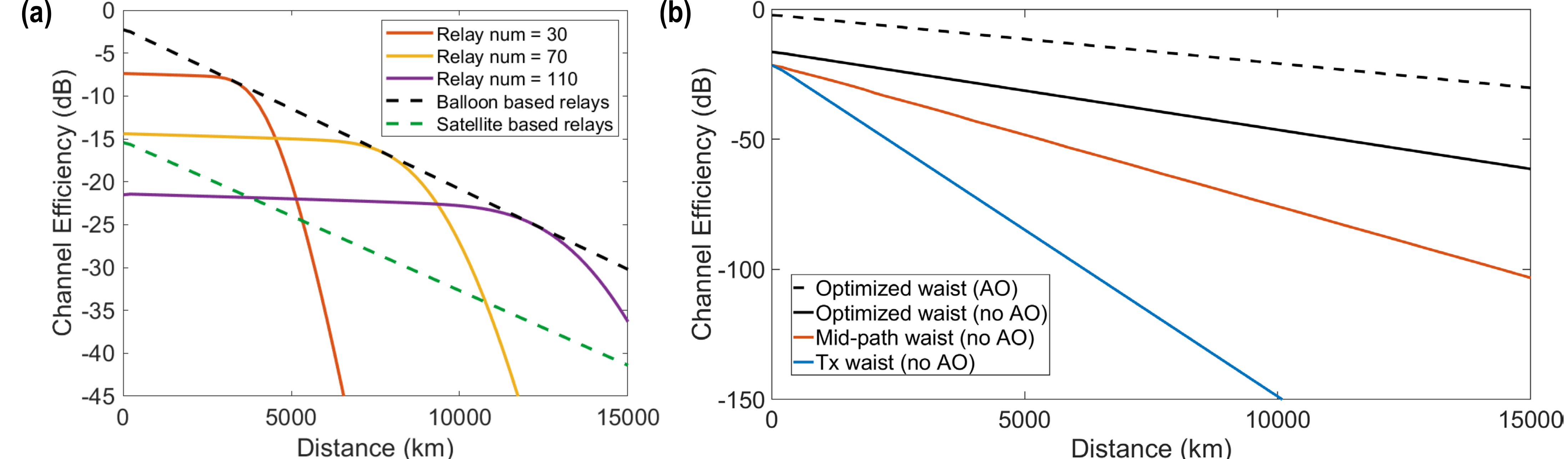}
    \caption{Channel efficiency of balloon-based relays as a function of the distance between servers, based on parameters listed in Appendix Table~\ref{tab:simulation_parameters}.
    \textbf{(a)} Optimized channel efficiency of balloon-based relays. Solid colored curves represent different relay balloon counts, showing that longer distances require more relay balloons (optimally spaced at $\sim$110 km). The black dashed line shows the maximum efficiency across all possible relay numbers, while the green dashed line shows the maximum efficiency achievable with satellite-based relays for comparison. We note that our simulation for satellite-based relays is more rigorous than that reported in Ref.~\cite{goswami_satellite-relayed_2023}, by further including wavefront distortion in both uplink and downlink channels (Appendix Section~\ref{Satellite}). \textbf{(b)} Comparison of maximum channel efficiency of balloon-based relays under different configurations. Optimized waist: Beam waist position optimized with (black dashed) and without (black solid) AO. Mid-path waist: Beam waist fixed at the midpoint of the path without AO (orange solid). Tx waist: Beam waist fixed at transmitter without AO (blue solid). At 10,000 km, our optimized configuration provides a 55 dB improvement over the typical mid-path waist configuration, establishing balloon-based relays as a practical backbone solution for global-scale quantum networking.
}      
\label{efficiency}
\end{figure*}

\begin{figure*}[t]
\centering
\includegraphics[width= 1 \linewidth]{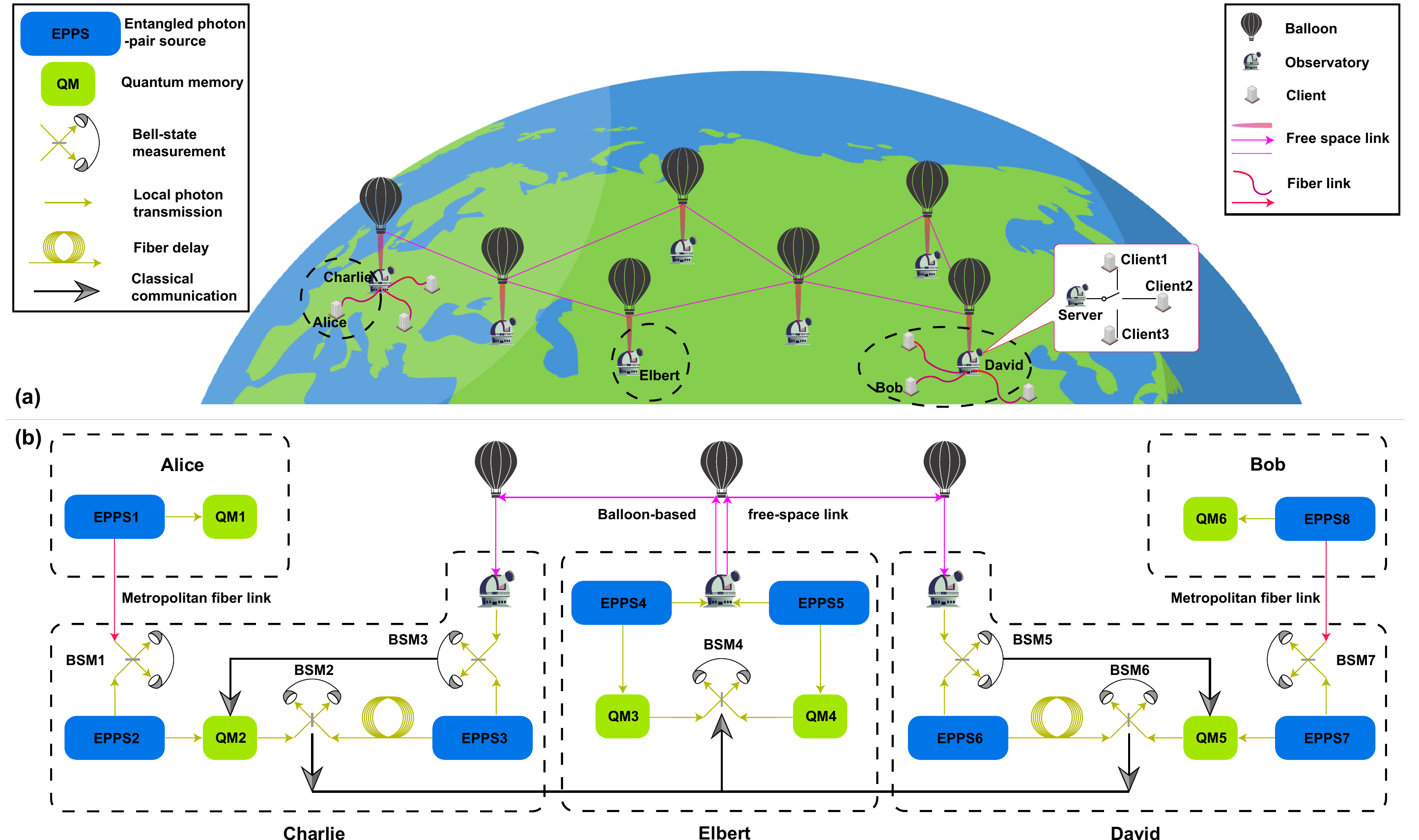}
\caption{The hybrid quadruple-link quantum repeater (H4QR) for global-scale entanglement distribution, with aerial relays and ground-based quantum repeaters. \textbf{(a)} Schematic representation of the connections between global-scale separated clients (Alice and Bob). Entanglement distribution between Alice and Bob involves four connections: two metropolitan fiber links (Alice $\leftrightarrow$ Charlie and Bob $\leftrightarrow$ David) and two balloon-based free-space relays (Charlie $\leftrightarrow$ Elbert and David $\leftrightarrow$ Elbert). \textbf{Inset:} One server serves multiple nearby clients via fiber links. These Clients can operate in sequence with fast optical switches or in parallel through wavelength multiplexing. \textbf{(b)} Detailed implementation of H4QR. Elementary entanglement is generated using Bell-state measurements (BSM1, BSM3, BSM5, BSM7) on photons from entangled photon pair sources (EPPSs). Quantum memories QM2 and QM5 temporarily store entangled photons until they are retrieved for synchronized entanglement swapping via BSM2 and BSM6. The final BSM4, performed on photons retrieved from QM3 and QM4, heralds the end-to-end entanglement between QM1 and QM6, completing the distribution process. 
}
\label{scheme fig}
%label要放在最下面，否则hyperref包会报错！
\end{figure*}

To enable entanglement distribution between globally separated clients, the proposed architecture requires only four long-distance connections (Fig.~\ref{scheme fig}): two metropolitan fiber links connecting clients to their local ground stations (Alice $\leftrightarrow$ Charlie, Bob $\leftrightarrow$ David), and two balloon-based free-space links between three ground stations (Charlie $\leftrightarrow$ Elbert, David $\leftrightarrow$ Elbert). We call this approach the hybrid quadruple-link quantum repeater (H4QR), designed for global-scale entanglement distribution between multiple distributed clients. 
The simple bilayer design of this hierarchical quantum repeater protocol leads to a relaxed requirement of quantum memories, resulting a simulated entanglement distribution rate (EDR) of sub-Hz over continental scale of 10,000 km. Importantly, all required parameters of quantum memories have been independently demonstrated in a unified physical system using Eu$^{3+}$:Y$_2$SiO$_5$ crystals (Table ~\ref{tab:repeater_parameters})~\cite{liu_nonlocal_2024,meng2025efficientintegratedquantummemory,ou2025multichannelhighdimensionalintegrated,lv2025minutescalephotonicquantummemory} and only probabilistic photon sources are needed, indicating the feasibility of implementing H4QR in the near future.

\section{Channel Efficiency of Balloon-based Relays}
\label{Channel Efficiency of Balloon-based Relays}
%在之前修改迭代的过程中，对于气球passive链路的描述渐渐消失了，审稿人就此提出了疑惑"It is not clear how the quantum signal is propagated and with which kind of (passive) optics"
We employ a chain of balloons hovering at 24 km as aerial platforms, which are equipped with acquisition, pointing, and tracking (APT) systems, telescopes with diameters of 0.6 m, and AO systems, functioning as passive optical relays. Quantum signals are sent from a local server to the nearest balloon, reflected and re-focused by balloons in the midway, and finally, collected and coupled into single-mode fiber (SMF) by the remote server's observatory. Our simulation incorporates all critical atmospheric effects including atmospheric attenuation, turbulence, scintillation,  wavefront aberrations and position jitter of balloons~\cite{andrews_laser_2005,scriminich_optimal_2022,PhysRevResearch.7.023199} alongside mechanical constraints such as pointing errors, with full parameters listed in Appendix Table~\ref{tab:simulation_parameters}. 

Fig.~\ref{efficiency}\textbf{(a)} shows the channel efficiency depending on different number of relay balloons. For a global-scale entanglement distribution spanning 20,000 km, approximately 92 relay balloons would be required between Charlie (David) and Elbert, resulting in an optimized channel efficiency of -21 dB at 10,000 km, which is 12 dB superior performance compared to satellite-based relays (Appendix Section~\ref{Satellite}). Fig.~\ref{efficiency}\textbf{(b)} reveals the efficiency gains through beam waist positioning and AO correction. In vacuum environment, a con-focal design of lens can eliminate the diffraction loss of Gaussian beam efficiently. However in atmosphere, the beam waist position in each relay section should be adjusted closer to the receiver for minimizing the influences of diffraction and beam wandering simultaneously (Appendix Section~\ref{subsubsection:Diffractive Properties of Gaussian-Beam Waves}). This configuration yields 30 dB and 100 dB enhancement in channel efficiency at a distance of 10,000 km, compared to the con-focal configuration and to that placing the beam waist at the transmitter, respectively. Phase aberrations, which leads to low collection efficiency to SMFs, can be minimized by employing a series of AO systems in the relay balloons (Appendix Section~\ref{subsubsection:fiber-coupling efficiency}). Implementation of AO further enhances the channel efficiency by 25 dB at a distance of 10,000 km. Taking the position jitter of balloons into consideration, we find the extra channel efficiency loss over 10000 km is within 0.5 dB, as discussed in Appendix Section~\ref{Unideal case of balloons}.
These optimizations establish balloon-based relays as a viable backbone channel for both quantum and classical communication networks. 

% which placing the beam waist at the midpoint of the propagation path
%Consequently, the channel losses in the horizontal links only leave atmospheric transmittance of -1.06 dB, receiver collection efficiency of -1.50 dB, fiber-coupling efficiency of -0.49 dB and optical loss of -16.14 dB, while the total loss in both downlink and uplink is about -1.8 dB. ***By comparing to simulations without any atmosphere, we conclude that our design has suppressed the ***90\% of the air disturbances despite operating near the Earth.*** 

\section{H4QR Protocol} 
To optimize network scalability and deployment feasibility, our design strategically locates all quantum repeater functionality at ground stations. Sophisticated, power-intensive quantum components, including quantum memories (QMs), entangled photon pair sources (EPPSs), and single-photon detectors (SPDs), are housed exclusively in ground stations, while the aerial platform employs standardized balloons equipped only with optical relays. Notably, one server can serve multiple local clients via fiber links (Inset in Fig.~\ref{scheme fig}a). Compared to other approaches relying on direct links between clients and aerial platforms~\cite{doi:10.1126/science.aan3211,boone_entanglement_2015,goswami_satellite-relayed_2023,gundogan_proposal_2021,gundogan_time-delayed_2024,meister_simulation_2025,chen_zero-added-loss_2023,wallnofer_simulating_2022}, our approach allows fast client switching and easy aerial-platform multiplexing, since aerial-platform telescope in our design only need to track the localized server, rather than multiple distributed clients.

To avoid the sensitivity to phase fluctuations in communication channels, all entanglement heralding and swapping operations in H4QR are based on Bell state measurements (BSMs) performed via two-photon detection~\cite{RevModPhys.83.33,liu_heralded_2021}. All BSMs are assumed to be implemented with standard linear optics elements, with an efficiency of 50\%~\cite{RevModPhys.83.33}. For metropolitan fiber connections, entanglement between QM1/QM6 (at Alice/Bob) and QM2/QM5 (at Charlie/David) is established through successful BSM1/BSM7 measurements. These BSMs jointly detect one photon from EPPS1/EPPS8 and another from EPPS2/EPPS7. Similarly, the balloon-based free-space links utilize BSM3/BSM6 to herald entanglement between one photon from EPPS3/EPPS6 at Charlie/David and QM3/QM4 at Elbert, achieved by projecting measurements on photons from EPPS4/EPPS5 with those from EPPS3/EPPS6. In the first stage, parallel entanglement generation occurs through independent BSM1, BSM7, BSM3, and BSM5 attempts until successful connections are established. In the second stage, photons stored in QM2/QM5 undergo prompt retrieval for BSM2/BSM6 implementation, synchronized via fiber delay lines to compensate for a short communication latency between BSM3(5) to local QM2(5). Successful BSM2 and BSM6 respectively herald entanglement between QM3-QM1 and QM4-QM6 pairs. Finally, end-to-end entanglement between QM1 and QM6 is heralded through BSM4 performed at Elbert on photons recalled from QM3 and QM4.

\begin{figure}[t] 
  \centering
  \includegraphics[width=1 \linewidth]{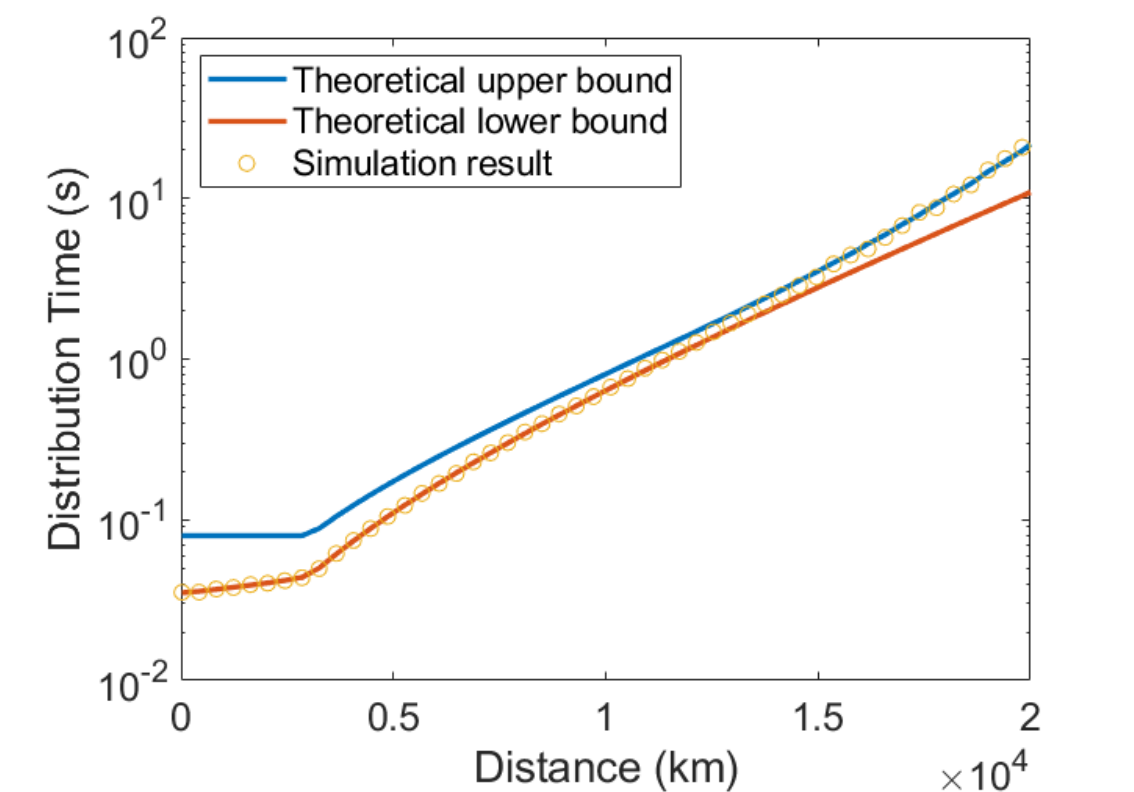}
  \caption{Theoretical and simulated entanglement distribution time of H4QR. The blue and red line represent the theoretical upper bound and the lower bound, respectively. The orange circles denote the numerical simulation based on Monte Carlo method. For short distances, the lower bound fits well with the simulation. For long distances, the simulated results getting closer to the upper bound.
  }
  \label{edr}  
\end{figure} 

\section{Distribution Time of H4QR}
\label{Distribution Time of H4QR}
Multiplexing is vital for boosting the EDR of quantum repeaters~\cite{collins_multiplexed_2007,simon_quantum_2007,PhysRevLett.113.053603}. It can be divided into two categories~\cite{collins_multiplexed_2007,simon_quantum_2007}: dependent modes requiring simultaneous retrieval of all modes, and independent modes permitting independent retrieval. We assume each QM has $n$ independent modes (for example, $n$ spatial channels) and $m$ dependent modes per independent mode (for example, $m$ temporal modes per spatial channel). Here we propose a theoretical method based on Markov chain to derive the upper bound of EDR, and follow the method provided in Ref.~\cite{collins_multiplexed_2007} to derive the lower bound of EDR, with details provided in Appendix Section~\ref{section:Calculations of Distribution time of H4QR}.
% It is challenging to derive an analytic solution for the EDR of multiplexed quantum repeaters, since variable number of QM modes could be occupied by residual entanglements.

The theoretical and simulation results for entanglement distribution time are plotted in Fig.~\ref{edr}, with the parameters of the QM listed in Table~\ref{tab:repeater_parameters}, and the free-space channel efficiency shown in Fig.~\ref{efficiency}. Other parameters include probabilistic EPPS with repetition rate $R=1$ MHz and emission probability $\rho=0.05$, metropolitan fiber length of 25 km between the client and the local server, with a loss coefficient of 0.18 dB/km and $\eta_\text{D}=0.9$. For distances up to 3,000 km, the distribution time is below 50 ms, primarily limited by the EPPS repetition rate and fiber losses. For global entanglement over 20,000 km, the distribution time is less than 20 s, primarily limited by free-space channel losses. 
%~\cite{Reddy:20}

\begin{figure}[t] 
  \centering
  \includegraphics[width=1 \linewidth]{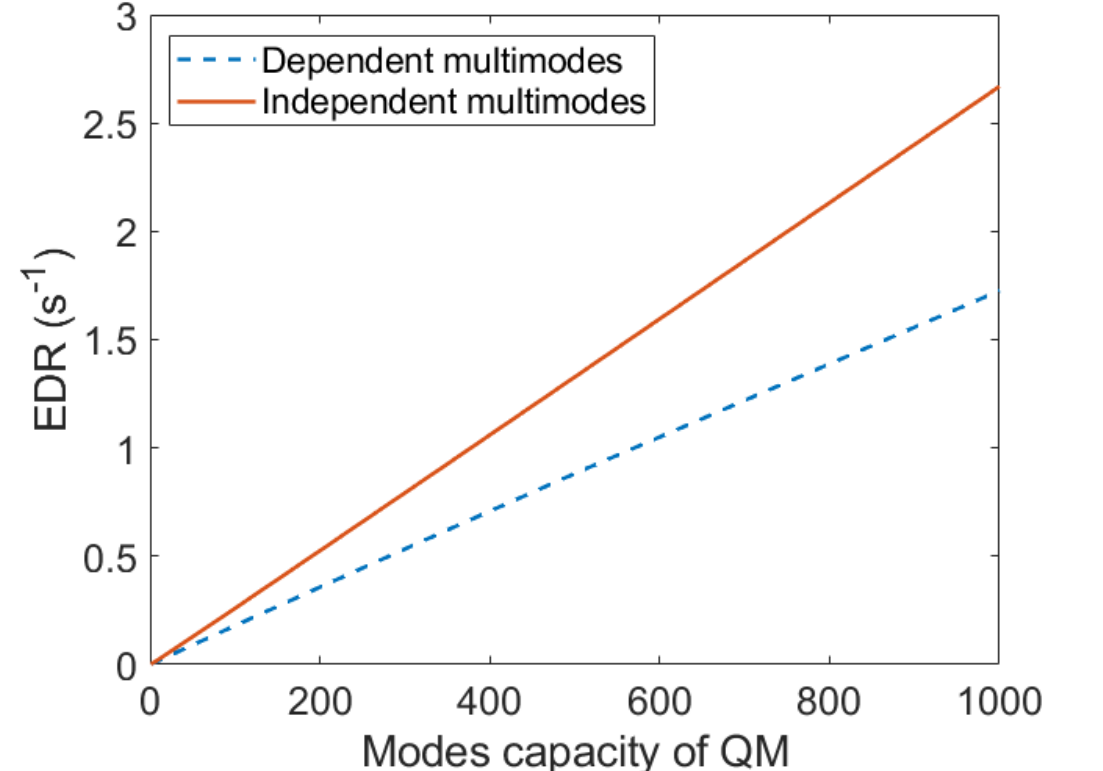}
  \caption{Theoretical upper bound of EDR as a function of the mode capacity of QMs at a distribution distance of $3000$ km. The blue dashed line: the horizontal axis represents the number of modes $m$, with $n=1$ fixed. The red solid line: the horizontal axis represents the number of modes $n$, with $m=1$ fixed.
  }
  \label{ind mode vs d mode}  
\end{figure}

\begin{table*}
    \centering
    \begin{ruledtabular}
    \begin{tabular}{cccccc}
        Parameter & Description &Value for calculation & Eu$^{3+}$:Y$_2$SiO$_5$ & Rb atomic ensemble* \\ 
        \hline
         
        $\eta_\text{M}$ &
        Efficiency&
        $80\%$&
        80.3\%~\cite{meng2025efficientintegratedquantummemory}&
        80.0\%~\cite{https://doi.org/10.1002/qute.202400480}\\

        %这篇文章写的太不自信了，让我很疑惑他们到底是不是最高效率
        %90.6\%~\cite{wang_efficient_2019}\\

        $\tau$ &
        Lifetime&
        $1$ s &
        27.6 s~\cite{lv2025minutescalephotonicquantummemory}&
        458 ms~\cite{PhysRevLett.126.090501}\\
        %存储寿命，冷原子和Pr均按照1/e时间给出
        
        $m$ &
       Number of dependent modes &
        $1000$&
        1097~\cite{liu_nonlocal_2024}&
        10~\cite{PhysRevLett.124.210504}\\
       %665~\cite{parniak_wavevector_2017}\\
        
        $n$ &
       Number of independent modes &
        $10$ &
        11~\cite{ou2025multichannelhighdimensionalintegrated}&
        225~\cite{pu_experimental_2017}\\
        %280~\cite{zhang_fast_2024}\\

    \end{tabular}
    \end{ruledtabular}
 
\caption{\label{tab:repeater_parameters} Key parameters of QMs assumed in the simulation of H4QR. For comparison, state-of-the-art demonstrations in Eu$^{3+}$:Y$_2$SiO$_5$ and Rb atomic ensemble are also listed. ``Independent modes" refers to the stored modes that can be retrieved individually. *Note that parameters for Rb atomic ensembles focus on emissive QMs, as absorptive QMs have so far been limited to short storage times.}
\end{table*}

H4QR achieves a fourfold higher EDR compared to satellite-based repeater schemes~\cite{boone_entanglement_2015}, despite the latter's reliance on more challenging techniques such as quantum nondemolition measurements.
Compared to fiber-based terrestrial quantum repeaters~\cite{RevModPhys.83.33}, H4QR consists of only 4 elementary links and 6 QMs, resulting in strong tolerance to the performances of QMs (Appendix Section~\ref{section EDR vs paras}).

\section{The Advantage of Independent Modes} Our rigorous analysis of the release process of multiplexed QMs in quantum repeaters reveals an inherent advantage for independent-mode multiplexing in boosting the EDR through faster entanglement swapping. To quantify this advantage, we define the average mode efficiency $\langle f_\kappa \rangle$ as the ratio of a QM mode's average operational duration to the total cycle time, where $\kappa$ denotes the number of modes ($\kappa=m$ for dependent modes and $\kappa=n$ for independent modes). In the regime of low entanglement generation probability, $\langle f_\kappa \rangle|_{\kappa=1}\approx\frac{2}{3}$, which recovers the characteristic factor of  $\frac{3}{2}$ in the EDR of hierarchical quantum repeater schemes~\cite{simon_quantum_2007,RevModPhys.83.33}.

In the case of multiplexing, $\langle f_n \rangle$ increases with $n$, while $\langle f_m \rangle$ decreases respectively with $m$, leading to a growing EDR gap between the two multiplexing strategies, as illustrated in Fig.~\ref{ind mode vs d mode} and Fig.~\ref{fig:mn}. This behavior stems from the fact that residual entanglement occupies fewer modes under independent storage and retrieval, which is theoretically proved in Appendix Section~\ref{section mn multip}.

In all hierarchical quantum repeater schemes, the total entanglement distribution time $T_{tot}$ scales as $\langle f_\kappa \rangle^{-N}$, where $N$ is the nesting level. For H4QR—a $N=1$ hierarchical scheme—independent multiplexing achieves an EDR gain of approximately 50\% as $\langle f_\kappa \rangle$ approaches unity. In multi-layer quantum repeater schemes which are typical for fiber channels, this advantage becomes substantially more pronounced, with independent multiplexing outperforming dependent multiplexing by a factor of roughly $(\frac{3}{2})^N$ in EDR.
 
%It is worth noting that independent modes significantly outperform dependent modes in multiplexing, achieving an additional improvement of more than 50\% in EDR, as shown in Fig.~\ref{ind mode vs d mode}. This is because the residual entanglements occupy less modes with independent storage and  retrieval, which is applicable to all multiplexed quantum repeater protocols~\cite{collins_multiplexed_2007,simon_quantum_2007,RevModPhys.83.33,PhysRevLett.113.053603}. Furthermore, the efficiency gap between two type of modes will become larger as the number of modes or nesting level grows, which is theoretically proved in Appendix Section~\ref{section mn multip}. 

%We further note that, when the distance between Charlie and David is less than 2000 km, it becomes unnecessary to conduct BSM4 at Elbert. In this scenario, one of the entangled photons from EPPS3 can be directly transmitted to David to execute BSM5 (no BSM3 required). This process establishes entanglement between one photon from EPPS3 and one photon from EPPS6. Consequently, the fiber delay line in Charlie should be substituted with a QM to await the outcomes of BSM5 and BSM6. The estimated EDR is close to (approximately 30\%) that of standard H4QR but it requires less QMs, EPPSs and BSMs, which could be easier to implement in practice.

\section{Technical Feasibility}
\label{sec5 Technical Feasibility}
Similar to other quantum repeater protocols, the performance of QMs is the most critical factor. These parameters of QM used in the simulation is chosen according to state-of-the-art experiments using Eu$^{3+}$:Y$_2$SiO$_5$ crystals, which we have demonstrated: 80.3\% storage efficiency using impedance-matched cavities~\cite{meng2025efficientintegratedquantummemory}, 27.6 s storage times under critical magnetic fields~\cite{lv2025minutescalephotonicquantummemory}, 11 independent spatial channels using 11 laser-written waveguides~\cite{ou2025multichannelhighdimensionalintegrated} and 1097 temporal modes for entanglement storage~\cite{liu_nonlocal_2024}. While these achievements were realized in separate setups, integrating them into a single system, i.e. a multichannel, cavity-enhanced, Eu$^{3+}$:Y$_2$SiO$_5$ QM operating at critical magnetic fields, is fundamentally feasible. Alternatively, emissive QMs based on Rb atomic ensembles~\cite{https://doi.org/10.1002/qute.202400480,PhysRevLett.126.090501,PhysRevLett.124.210504,pu_experimental_2017,liu_creation_2024} could replace the absorptive QM and EPPS in each H4QR nodes, offering comparable performance (see Table~\ref{tab:repeater_parameters}). %Recent advances have also demonstrated addressing multiple distinguishable single-particle emitters within a single device~\cite{doi:10.1126/science.abc7821,doi:10.1126/science.ado6471,PRXQuantum.5.020308,ruskuc_multiplexed_2025,li2025parallelizedtelecomquantumnetworking}, but scaling such systems to a large number of modes remains a significant challenge.

H4QR is compatible with a wide range of EPPS, including both probabilistic sources like spontaneous parametric down-conversion (SPDC)~\cite{liu_heralded_2021,meng2025efficientintegratedquantummemory} and deterministic sources such as single atoms, ions, or quantum dots~\cite{tang_storage_2015}. Our simulation assumes the use of mature SPDC-based sources, which have already enabled quantum storage in Eu$^{3+}$:Y$_2$SiO$_5$~\cite{liu_nonlocal_2024,meng2025efficientintegratedquantummemory}.  SPDC sources allow independent wavelength selection for each photon: one photon operates at 1537 nm for long-distance transmission, while the other operates at 580 nm for long-lived storage in Eu$^{3+}$:Y$_2$SiO$_5$. The SPDC sources emit with a repetition rate of $R=1$ MHz, matching the typical bandwidth of long-lived Eu$^{3+}$:Y$_2$SiO$_5$ QMs~\cite{ma_one-hour_2021}. The current simulation assumes no multiplexing in EPPS, with photons delivered sequentially to specific spatial and temporal modes of the QMs. Frequency-multiplexed SPDC sources could further enhance EDR by enabling parallel writing into frequency-multiplexed QMs~\cite{PhysRevLett.113.053603,PhysRevLett.123.080502}.

% This EPPS configuration leverages cavity-enhanced SPDC technology, which is well-established~\cite{meng2025efficientintegratedquantummemory}.
%***Discussions on satellite relays**

%Balloon-based communication is feasible since quantum key distribution between balloons and the ground~\cite{wang_direct_2013} and optical communication between balloons~\cite{hemmati_demonstration_2017} have been demonstrated. 

Although station-keeping for stratospheric balloons represents a technical challenge, a reinforcement learning–based flight controller developed by Google Research has successfully maintained superpressure balloons in the Loon fleet within 50 km of their target locations for 39 days \cite{bellemare_autonomous_2020}. More recent work has further reduced this deviation to under 10 km over 24 hours ~\cite{XU2022733}.
Besides, other high-altitude platforms—such as vented solar balloons~\cite{10116024} and solar-powered drones~\cite{9450597}—also represent viable candidates for implementing the proposed relay system.

%~\cite{Lally1967SuperpressureBF} 
%Furthermore, recent researches demonstrate more energy-efficient and precise scheme for balloon navigation~\cite{11068473,brown_stratospheric_2025,electronics14010204,11068667,electronics13204032}, No-Fly Zone avoidance~\cite{10670807} and launch site optimization~\cite{10802275}, which lay a solid foundation for our protocol.  

%However, the long-term operating costs and stability of balloons still represent technical challenges, despite rapid progress in these areas. 
%Meanwhile, the satellite-based relay~\cite{goswami_satellite-relayed_2023} is another potential backbone channel option for H4QR (Appendix Section~\ref{Satellite}), offering comparable channel efficiency and leveraging well-developed satellite communication technology.

\section{Discussion}
In summary, we establish balloon-based aerial relays as a novel backbone channel for quantum networks and introduce the H4QR protocol, which further integrates ground-based quantum repeaters to enable high-speed global-scale entanglement distribution. The core strength of H4QR lies in its use of QMs that all critical parameters have been individually validated in a unified physical system using Eu$^{3+}$:Y$_2$SiO$_5$ crystals, paired with
EPPSs leveraging well-established SPDC technology. Successful implementation depends on three key milestones: (1) establishing low-loss aerial relays, (2) consolidating all QM parameters into a single experimental setup and (3) deploying hybrid-channel quantum repeaters. Additionally, the Eu$^{3+}$:Y$_2$SiO$_5$ crystal platform demonstrates remarkable potential for developing transportable quantum memories~\cite{lv2025minutescalephotonicquantummemory,ma_one-hour_2021,Bland-Hawthorn:21,gundogan_time-delayed_2024}, paving the way for advanced quantum networks that seamlessly combine aerial relays, optical fibers, and mobile end nodes. Beyond quantum applications, the development of global-scale low-loss aerial relays could also unlock new opportunities for classical communication systems.

\begin{acknowledgments}
P.-X. L. and Y.-P. L. contributed equally to this work. This work is supported by the National Natural Science Foundation of China (Nos. 12222411 and 11821404) and the Quantum Science and Technology-National Science and Technology Major Project (No. 2021ZD0301200). Z.-Q.Z acknowledges the support from the Youth Innovation Promotion Association CAS.
\end{acknowledgments}

\section*{Data Availability}
The data that support the findings of this article are openly available~\cite{dataset}

\appendix
\section{Simulation Methods of Free-Space Propagation}
\label{channeleff simulation}
In this section, we first outline the geometrical parameters employed in our calculations (Section~\ref{subsection:Geometrical Considerations}). We then provide an overview of the factors influencing channel efficiency, followed by a detailed explanation of the calculation process (Section~\ref{subsection:Channel Efficiency in Free Space}). The formulas presented primarily draw from Ref.~\cite{andrews_laser_2005, scriminich_optimal_2022}, with two key innovations in our approach: (1) optimization of beam waist positions (Section~\ref{subsubsection:Diffractive Properties of Gaussian-Beam Waves}) and (2) correction of accumulated phase aberrations using a series of adaptive optics (AO) systems in relay balloons (Section~\ref{subsubsection:fiber-coupling efficiency}). These two enhancements were critical in achieving a practical level of channel efficiency. Subsequently, we perform a parameter exploration and summarize the values used in our final calculations for the balloon-based configuration (Section~\ref{Parameter Exploration}). For simplicity, our simulations assume that all balloons are evenly distributed, but pointing error due to positional jitter is taken into account. A brief discussion of deviations from this idealized scenario is provided (Section~\ref{Unideal case of balloons}). Finally, we detail the simulation of channel efficiency for satellite-based relays (Section~\ref{Satellite}).
 
\subsection{Geometrical Considerations} 
\label{subsection:Geometrical Considerations}
\begin{figure*}[tb]
\centering
\includegraphics[width= 1.0 \linewidth]{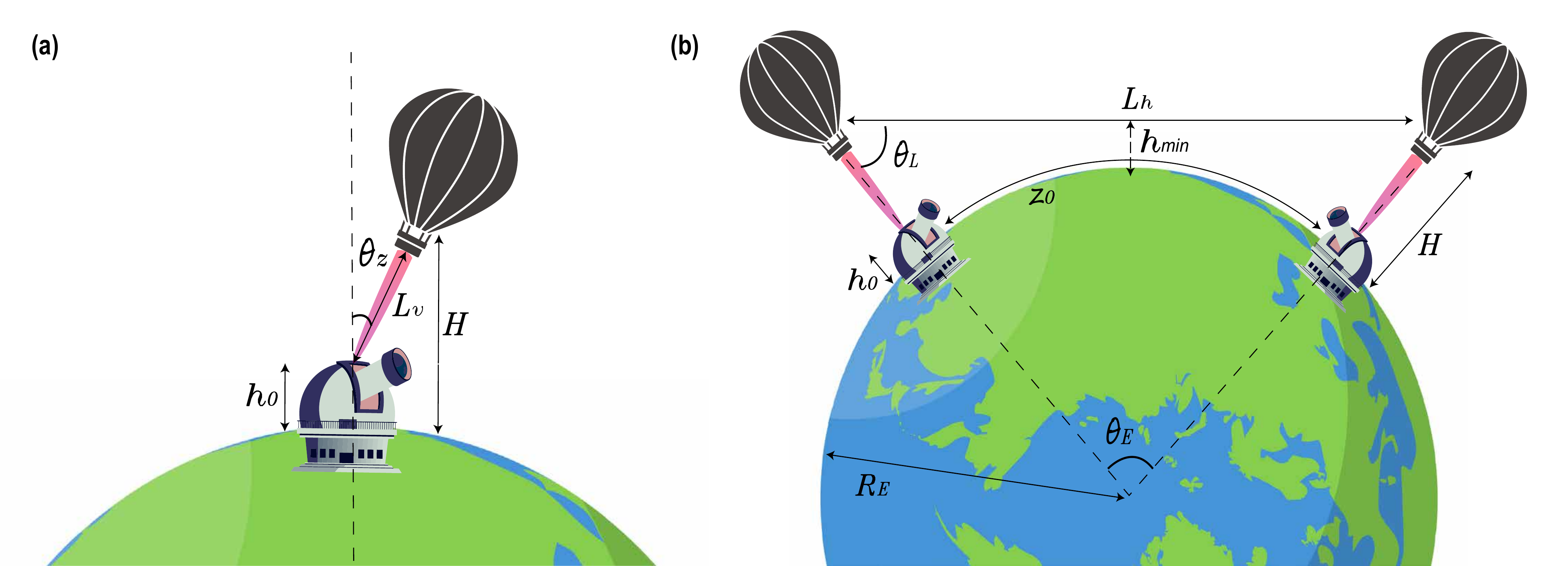}
\caption{ 
Representation of the geometrical parameters used in our simulation. Balloons are positioned at an altitude $H$, and ground station transmitters are located at an altitude $h_0$. \textbf{(a)} In the uplink and downlink channels, $L_v$ is the light propagation distance (Eq.\ref{eq.L_v}). $\theta_z$ is the zenith angle, which is set to zero in our simulation as the balloons are assumed to float directly above the observatories of servers. \textbf{(b)} In the horizontal channel, $L_h$ is the light propagation distance (Eq.\ref{eq.L_h}), $R_E$ is the radius of the Earth, $z_0$ is the arc length between two ground stations, $\theta_E$ is the corresponding subtending angle (Eq.\ref{eq.thetaE}), $h_\text{{min}}$ is the minimum height along the path between the two balloons (Eq.\ref{eq.hmin}). For all relay balloons, light is not coupled into single-mode fibers (SMFs) but corrected by AO systems to reduce channel loss.
}
\label{figS1}
\end{figure*}

The geometrical parameters used for beam propagation in free space are illustrated in Fig.~\ref{figS1}. In our analysis, we assume that the Earth is a perfect sphere with a radius of $R_E$ = 6,371 km. 
For uplink and downlink channels, the propagation distance $L_v$ is given by
\begin{equation}
    L_v=(H-h_0)\sec(\theta_z).
\label{eq.L_v}
\end{equation}
For horizontal channel, we assume that all balloons are distributed along the great circle of the Earth for simplicity:
\begin{equation}
\theta_E = \frac{z_0}{R_E},
\label{eq.thetaE}
\end{equation}
\begin{equation}
L_h = 2(R_E + H)\sin\left(\frac{\theta_E}{2}\right), 
\label{eq.L_h}
\end{equation}
\begin{equation}
    h_{\text{min}} = (R_E + H)\cos\left(\frac{\theta_E}{2}\right) - R_E.
\label{eq.hmin}
\end{equation}
If the separation distance 
$z_0$ between ground stations becomes too large, the signal path may be obstructed by the Earth's curvature, leading to $h_{\text{min}}<0$. In such cases, this issue can be mitigated by either increasing the altitude of the balloons or introducing additional relay nodes.
%%%%%%%%%%%%%%%%%%%%%%%%%%%%%%%%%%%%%%%%%%%%%%% 
%Fig. 1 %%%%%%%%%%%%%%%%%%%%%%%%%%%%%%%%%%
%%%%%%%%%%%%%%%%%%%%%%%%%%%%%%%%%%%%%%%%%%%%%%%%%%%%%%%%
%%%%%%%%%%%%%%%%%%%%%%%%%%%%%%%%%
\subsection{Channel Efficiency in Free Space} 
\label{subsection:Channel Efficiency in Free Space} 

In our scenario, the whole channel consists of an uplink, a downlink and several horizontal channels between relay balloons. Noting that the optical beam coupled into the SMF only at the final stage, i.e. in the downlink channel. Therefore, the total channel efficiency can be expressed as
\begin{equation}
\eta_\text{tol}=\eta_\text{uplink}\cdot\eta_\text{downlink}\cdot\eta^{N-1}_\text{horizontal}\cdot\eta_{\text{SMF}},
\end{equation}
where the fiber-coupling efficiency $\eta_{\text{SMF}}$ is considered separately based on the overall channel characteristics. Here, $N$ denotes the number of relay balloons, dividing the horizontal path into $(N-1)$ segments, and $\eta_\text{uplink}, \eta_\text{downlink}, \eta_\text{horizontal}$ are channel efficiencies given by 
\begin{equation}
    \eta_{\text{CH}}=\eta_{t}\eta_{r}\cdot\eta_{atm} \cdot\eta_{D_{R_x}},
\end{equation}
where $\eta_{t}$ ($\eta_{r}$) is the optical transmittance of the transmitter (receiver), $\eta_{\text{atm}}$ is the atmospheric transmittance, and $\eta_{D_{R_x}}$ is the receiver collection efficiency.
%%%%%%%%%%%%%%%%%%%%%%%%%%%%%%%%%%%%%%%%%%%%%%%%%%%%%%%%%%%%%%%%%%%%%

%%%%%%%%%%%%%%%%%%%%%%%%%%%%%%%%%%%%%%%%%%%%%%%%%%%%%%%%%%%%%%%%%%%%%%%
\subsubsection{Atmospheric Transmittance}
When propagating through the atmosphere, the light intensity $I$ attenuates as 
\begin{equation}
    \diff I=-\alpha(h(x))I\diff x\ (0\leq x\leq L_0),
    \label{equa1}
\end{equation}
where $L_0$ uniformly denotes the propagation distance from transmitter to receiver, i.e. $L_0=L_v$ for the uplink/downlink channel and $L_0=L_h$ for the horizontal channel. We obtain the attenuation coefficient $\alpha(h)$ using the LOWTRAN software package (Ref.~\cite{1988ugls.rept.....K}), %and in particular the MATLAB code \cite{Noah2025LOWTRAN7}
where $h=h(x)$ represents the altitude at each point along the path. For uplink and downlink channels, it is given by
\begin{equation} 
    h(x)=x\cos{\theta_z},
\end{equation}
and for horizontal channel,
\begin{equation}
    h(x)=\sqrt{(R_E+H)^2+x^2-2(R_E+H)x\cos(\theta_L)}-R_E,
\end{equation}
where $\cos(\theta_L)=\frac{L_h}{2(R_E+H)}$, as illustrated in Fig.~\ref{figS1}. Substituting into Eq.\ref{equa1}, we can integrate the attenuation to get the atmospheric transmittance as
%\begin{equation}
    %\diff \ln{I}=-\alpha(h(x))\diff x\ (0\leq x\leq L),\
%    \ln{(I/I_0)}=-\int_{0}^{L_0} \alpha(h(x))\diff x.
%\end{equation}
% Hence, the atmospheric transmittance is given by
\begin{equation}
    \eta_{atm}=I/I_0=\exp{\left(-\int_{0}^{L_0} \alpha(h(x))\diff x\right)}.
\end{equation}           
%%%%%%%%%%%%%%%%%%%%%%%%%%%%%%%%%%%%%%%%%%%%%%%%%%%

%%%%%%%%%%%%%%%%%%%%%%%%%%%%%%%%%%%%%%%%%%%%%%%%%%%%%%%%
\subsubsection{Diffractive Properties of Gaussian-Beam Waves}
\label{subsubsection:Diffractive Properties of Gaussian-Beam Waves}
According to Ref.~\cite{andrews_laser_2005}, we define the %input plane 
transmitter beam parameters for subsequent use as
\begin{equation}
    \Theta_0=1-\frac{L_0}{F_0},\ \bar{\Theta}_0=1-\Theta_0,\ \Lambda_0=\frac{2L_0}{kW_0^2},
\end{equation}
where $k=2\pi/\lambda$ is the wave number corresponding to the wavelength $\lambda$, $W_0$ and $F_0$ denote the beam radius and curvature radius at the transmitter, respectively. The corresponding %output plane 
receiver beam parameters are given by
\begin{equation}
   \Theta=\frac{\Theta_0}{\Theta_0^2+\Lambda_0^2},\ \bar{\Theta}=1-\Theta,\ \Lambda=\frac{\Lambda_0}{\Theta_0^2+\Lambda_0^2}.
\end{equation}
In our simulation, $W_0$ is not necessarily the beam waist, but rather the optimal initial beam radius that maximizes the overall channel efficiency. Hence, the diffraction-limited beam radius at the receiver should be expressed in a more general form 
\begin{equation}
    W_{\text{diff}}=W_0\sqrt{\Theta_0^2+\Lambda_0^2}. 
\end{equation}
Given the beam radius $W_0$ at the transmitter, the potential distance between the transmitter and the beam waist, denoted by $d$, lies within the range $d\in\left[0, \frac{\pi W_0^2}{2\lambda}\right]$. By traversing all possible values of $d$, we can determine the optimal waist position and the corresponding minimum channel loss. According to our simulation results, when the propagation distance falls within the interval $L_0\in\left[0, \frac{\pi W_0^2}{2\lambda}\right]$, the optimal beam waist is positioned within the range of approximately $0.7L_0$ to $0.8L_0$ (i.e. $d\in[0.7L_0, 0.8L_0]$). As shown in Fig.~\ref{efficiency}\textbf{(b)}, without the AO systems, our configuration results in 30 dB efficiency enhancement at the distance of 10,000 km compared to placing the beam waist at the midpoint of the propagation path $(d=L_0/2)$, and 100 dB enhancement relative to positioning it at the transmitter $(d=0)$, respectively.
%%%%%%%%%%%%%%%%%%%%%%%%%%%%%%%%%%%%%%%%%%%%%%%%%%%%%%%%%%%%%%%%%
%%%%%%%%%%%%%%%%%%%%%%%%%%%%%%%%%%%%%%%%
\subsubsection{Collection Efficiency}
\label{section:Collection Efficiency}
According to the Kolmogorov’s theory of turbulence (Ref.~\cite{kolmogorov1991local, andrews_laser_2005}), %寻找原始文献出处
considering the turbulence-induced beam broadening and wandering, the average and instantaneous beam size at receiver can be represented by long-term ($W_{\text{LT}}$) and short-term ($W_{\text{ST}}$) beam radius, respectively.
\begin{equation}
    W_{\text{LT}}=W_{\text{diff}}\sqrt{1+T}, 
\end{equation}
\begin{equation}
    W_{\text{ST}}=\sqrt{W^2_{\text{LT}}-\langle r_c^2 \rangle}.
\end{equation}
For uplink and downlink channels:
\begin{equation}
   T=4.35\mu_{2}\Lambda^{5/6}k^{7/6}(H-h_0)^{5/6}\sec^{11/6}(\theta_z),
\end{equation}

\begin{widetext}
\begin{equation}
\begin{aligned}
\langle r_c^2 \rangle=7.25(H-h_0)^2\sec^{3}(\theta_z)W_0^{-1/3}\int_{h_0}^HC_n^2(h)\xi^2(h)\left[\frac{1}{(\Theta_0+\bar{\Theta}_0\xi(h))^2+1.63\sigma_R^{12/5}\Lambda_0(1-\xi(h))^{16/5}}\right]^{1/6}\diff h,
\end{aligned}
\end{equation}
\end{widetext}

where $\xi(h)$ is a normalized distance variable introduced in the path amplitude ratio for Gaussian beam propagation, defined as
\begin{equation}
    \xi(h)=\left\{
\begin{aligned}
\frac{h-h_0}{H-h_0},\ \text{for downlink channel} \\
1-\frac{h-h_0}{H-h_0},\ \text{for uplink channel}
\end{aligned}
\right.
\end{equation}
\begin{equation}
    \mu_2=\int_{h_0}^HC_n^2(h)\xi^{5/3}(h)\diff h,
\end{equation}
$\sigma_R^2$ is Rytov variance, given by
\begin{equation}
    \sigma_R^2=2.25(H-h_0)^{5/6}k^{7/6}\sec^{11/6}(\theta_z)\int_{h_0}^HC_n^2(h)\xi^{5/6}(h)\diff h,
\end{equation}
and $C_n^2(h)$ is the refractive index structure constant (Ref.~\cite{Hufnagel:64, NAP28728, hufnagel1978propagation}). We use the Hufnagel-Valley 5/7 (H-V$_{5/7}$) model in our simulation (Ref.~\cite{Valley:80, ulrich1988hufnagel, andrews_laser_2005}).   
For horizontal channel:  
\begin{equation}
    T=1.63(\sigma_R^2)^{6/5}\Lambda,
\end{equation}

%\begin{widetext}
\begin{equation}
%\begin{aligned}
\langle r_c^2 \rangle=7.25C_n^2L_h^{3}W_0^{-1/3}\int_{0}^1\xi^2\left[\frac{1}{(\Theta_0+\bar{\Theta}_0\xi)^2+1.63\sigma_R^{12/5}\Lambda_0(1-\xi)^{16/5}}\right]^{1/6}\diff \xi,
%\end{aligned}
\end{equation}
%\end{widetext}

with
\begin{equation}
 \xi(x)=1-\frac{x}{L_h},(\ 0\leq x \leq L_h),
\end{equation}
\begin{equation}
 \sigma_R^2=1.23C_n^2k^{7/6}L_h^{11/6}.
\end{equation}
Note that for the horizontal channel, we take $C_n^2=C_n^2(h_\text{min})$, which gives a lower bound for the total channel efficiency.

By employing the acquisition, pointing, and tracking (APT) system, beam wandering can be effectively suppressed, making the collection efficiency primarily dependent on the instantaneous beam size, i.e. the short-term beam radius. However, the APT system inevitably introduces latency between the emission and reception of the beacon light. During this delay, the balloons drift with the wind in random directions, resulting in a misalignment between the transmitter and receiver and consequently increasing the pointing error. Assuming the balloons drift at the average transverse wind speed, $\bar{v}_\text{wind}$, the pointing error can be modeled as an additional contribution to the beam broadening variance:
\begin{equation}
    W^2=W_{\text{ST}}^2+(\bar{v}_\text{wind}\cdot t_0)^2,
\end{equation}
where $t_0$ is the system latency, given by $t_0=L_0/c$ with $c$ being the speed of light. Specifically, the pointing error is found to contribute approximately 0.09\% to the total channel loss at a distance of 10,000 km.
 
%The pointing error can be eliminated by APT system, since the balloons remain almost stationary during once signal transmission, compared with satellites~\cite{PhysRevResearch.7.023199}.
%The pointing error is assumed to be negligible, as the balloons remain stationary above the observatories.
Finally, the collection efficiency at the receiver, $\eta_{D_{R_x}}$, is given by
\begin{equation}
\label{equalcollection}
    \eta_{D_{R_x}}=1-\exp \left(-\frac{D^2_{R_x}}{2W^2}\right),
\end{equation}
which describes the fraction of a Gaussian beam with beam-spot radius $W$ collected by a receiver with a circular aperture of diameter $D_{R_x}$.
%%%%%%%%%%%%%%%%%%%%%%%%%%%%%%%%%%%%%%%%%%%%%%%%%%%%%%

%%%%%%%%%%%%%%%%%%%%%%%%%%%%%%%%%%%%%%%%%%%%%%%%%%%%%%%%
\subsubsection{Fiber-coupling Efficiency}
\label{subsubsection:fiber-coupling efficiency}
When propagating through a turbulent atmosphere, the optical field will be distorted. According to Rytov theory (Ref.~\cite{1989psr4.book.....R, tatarskii1971effects, andrews_laser_2005}), the perturbed field can be described as %寻找原始文献 
\begin{equation}
    \Psi(\Vec{r},t)=\Psi_0(\vec{r},t)\exp{\left[\chi(\vec{r},t)+i\phi(\vec{r},t)\right]},
\end{equation}
where $\Psi_0(\vec{r},t)$ is the unperturbed optical field, $\phi(\vec{r},t)$ is the phase aberrations that can be partially corrected by adaptive optics (AO), and $\chi(\vec{r},t)$ is the log-amplitude fluctuations associated with scintillation.

As discussed in Ref.~\cite{Canuet:18}, the matching between the unperturbed incident beam $\Psi_0(\Vec{r},t)$ and the SMF mode can be evaluated by
\begin{equation}
    \Omega_0=\frac{\braket{\Psi_0|M_0}_P}{\left[\braket{\Psi_0|\Psi_0}_P\times\braket{M_0|M_0}_P\right]^{1/2}},
\end{equation}
where $M_0$ denotes the SMF mode in the plane of telescope pupil, $P$ denotes the pupil transmittance as
\begin{equation}
P(\Vec{r})=\begin{cases}
1, & 0 \leq \frac{2|\vec{r}|}{D_{R_x}} \leq 1\\
0, & \text{otherwise}
\end{cases} 
\end{equation}
and the operator $\braket{\cdot|\cdot}_Z$ refers to the scalar product with weight $Z$ as
\begin{equation}
\braket{X|Y}_Z  \triangleq \iint Z(\vec{r}) \cdot X(\vec{r}) \cdot Y^*(\vec{r}) \, d^2\vec{r}.
\end{equation}
Then the coupling efficiency of unperturbed beam can be represented by
\begin{equation}
    \eta_0=|\Omega_0|^{2}.
\end{equation}

Similarly, the matching between the distorted incident beam and SMF mode and the coupling efficiency can be written as
\begin{equation}
    \Omega=\frac{\braket{\Psi|M_0}_P}{\left[\braket{\Psi|\Psi}_P\times\braket{M_0|M_0}_P\right]^{1/2}}, 
\end{equation}
\begin{equation}
\eta_\text{SMF}=|\Omega|^{2}.
\end{equation}
We can get a more concise expression by the ratio 
\begin{equation}
\frac{\eta_\text{SMF}}{\eta_0}=\left|\frac{\Omega}{\Omega_0}\right|^2=\left|\frac{\braket{\Psi|M_0}_P}{\braket{\Psi_0|M_0}_P}\right|^2. 
\end{equation}
           
Assuming that the phase aberrations and log-amplitude fluctuations are statistically independent (Ref.~\cite{Fried:66}), the coupling efficiency $\eta_{\text{SMF}}$ can be written as
\begin{equation}
    \eta_{\text{SMF}}=\eta_0\cdot\eta_{\text{AO}}\cdot\eta_{s},
\end{equation}
where $\eta_{\text{AO}}$ is the coupling efficiency due to wavefront aberrations with AO corrections, and $\eta_{s}$ is the coupling efficiency due to atmospheric scintillation.

\paragraph{$\eta_0$}
According to Ref.~\cite{Ruilier:01, scriminich_optimal_2022}, with no central obscuration (by using off-axis telescopes) and the best matching between the unperturbed beam and the MFD of the SMF, we set $\eta_0$ to its maximum 81.5\%. 
%It should be noted that this term is only taken into account when light is coupled into the SMF. 
%For those relay balloons which do not use SMF for collection, $\eta_0$ should be set at $1$, in uplink and all horizontal links in our architecture. 
 
\paragraph{$\eta_{\text{AO}}$}
\label{AOefficiency}
According to Ref.~\cite{Canuet:18}, the instantaneous matching between the phase-aberration-only fluctuations where $\chi(\Vec{r},t)=0$, $\Psi_{\Phi}=\Psi_0\exp{\left[i \phi(\vec{r},t)\right]}$ and the SMF mode is given by 

\begin{widetext}

\begin{equation}
\frac{\Omega_\Phi}{\Omega_0}=\frac{\braket{\Psi|M_0}_P}{\braket{\Psi_0|M_0}_P}=\braket{\exp(i \phi)}_{W_P}\simeq\exp{\left(i\braket{\phi}_{W_P}\right)}\exp{\left(-\frac{\sigma^2_{W_P}(\phi)}{2}\right)},
\end{equation}
\end{widetext}

where $W_P=PM_0$ and $\sigma^2_{W_P}(\phi)$ represents the spatial variance of the phase aberrations over the pupil. The instantaneous coupling efficiency is approximated as 
\begin{equation}
    \eta_\Phi\simeq\eta_0 \exp{\left[-\sigma^2_{W_P}(\phi)\right]}=\eta_0 \exp{\left[-\sum_{\mathrm{n},\mathrm{m}}(b^{\mathrm{m}}_{\mathrm{n}})^2\right]},
\end{equation}
where $b^{\mathrm{m}}_{\mathrm{n}}$ corresponds to the Zernike coefficient of radial degree $\mathrm{n}$ and azimuthal degree $\mathrm{m}$ as
\begin{equation} 
    \phi(\Vec{r},t)=\sum_{\mathrm{n},\mathrm{m}}b^{\mathrm{m}}_{\mathrm{n}}(t)Z^{\mathrm{m}}_{\mathrm{n}}(\Vec{r}).
\end{equation} 
Here, the radial degree $\mathrm{n}$ and azimuthal degree $\mathrm{m}$ differ from the independent modes $n$ and dependent modes $m$ mentioned in Section~\ref{section:Calculations of Distribution time of H4QR}. After being corrected by the AO system with limited bandwidth and maximum corrected radial order $N_\text{AO}$, the instantaneous coupling efficiency turns to 
\begin{equation} 
    \eta_\text{AO} = \frac{\eta_\Phi}{\eta_0}\simeq \exp{\left[-\sum_{\mathrm{n},\mathrm{m}}\gamma_{\mathrm{n}}^2(b^{\mathrm{m}}_{\mathrm{n}})^2\right]},
\end{equation}
where $\gamma_{\mathrm{n}}^2$ is the mode attenuation factor for the $n$th order aberration coefficients.  For $\mathrm{n}>N_\text{AO}$, $\gamma_{\mathrm{n}}^2 = 1$. Otherwise, it can be expressed as the ratio of the phase variance after correction to that before correction, which can be derived from Parseval’s theorem (Ref.~\cite{conan1995wavefront, roddier2004adaptive, scriminich_optimal_2022}):
\begin{equation}
    \gamma_{\mathrm{n}}^2 = \frac{\int |W_{\mathrm{n}}(\tilde{\nu})|^2 |\epsilon(\tilde{\nu})|^2 \, d\tilde{\nu}}{\int |W_{\mathrm{n}}(\tilde{\nu})|^2 \, d\tilde{\nu}},
    \label{eq:gamma}
\end{equation}
where $|W_{\mathrm{n}}(\tilde{\nu})|^2$ represents the power spectral density of the ${\mathrm{n}}$-th order wavefront phase fluctuations, $\tilde{\nu}$ is the temporal frequency of the AO loop, and $\epsilon(\tilde{\nu})$ represents the transfer function between the residual phase and the turbulent wavefront phase fluctuations which is related to the AO system’s open-loop transfer function $G(\tilde{\nu})$:
\begin{equation}
    \epsilon(\tilde{\nu}) = \frac{1}{1 + G(\tilde{\nu})}.
    \label{eq:epsilon}
\end{equation} 
Considering a typical AO system with a pure integrator based on a Shack--Hartmann wavefront sensor, and correction with a deformable mirror, the open-loop transfer function is then given by
\begin{equation}
    G(\tilde{\nu}) = K_I \frac{e^{-\tau_{AO} \tilde{\nu}} \left(1 - e^{-T_{AO}\tilde{\nu}}\right)}{(T_{AO} \tilde{\nu})^2},
    \label{eq:G}
\end{equation}
where $K_I$ is the gain of the integrator, $\tau_{AO}$ is the overall delay of the control-actuator stage, and $T_{AO}$ is the wavefront sensor integration time. All these numerical parameters are listed in Table~\ref{tab:simulation_parameters}.

As for the power spectral density, $|W_{\mathrm{n}}(\tilde{\nu})|^2$, it scales polynomially with a cut-off frequency $\tilde{\nu}_c^{({\mathrm{n}})}$:
\begin{equation}
    |W_{\mathrm{n}}(\tilde{\nu})|^2 \sim 
    \begin{cases}
        \tilde{\nu}^{-2/3}, & \tilde{\nu} \leq \tilde{\nu}_c,\; {\mathrm{n}}=1 \\
        \tilde{\nu}^{0}, & \tilde{\nu} \leq \tilde{\nu}_c,\; {\mathrm{n}} \neq 1 \\
        \tilde{\nu}^{-17/3}, & \tilde{\nu} > \tilde{\nu}_c
    \end{cases}
    \label{eq:Wn}
\end{equation}
with
\begin{equation}
    \tilde{\nu}_c^{({\mathrm{n}})} \approx 0.3({\mathrm{n}}+1)\bar{v}_\text{wind}/D_{{R_x}}.
    \label{eq:cutoff}
\end{equation}

Noting that in our scheme, the beam is coupled to the SMF only in the last channel, i.e. the downlink. Thus, the overall phase aberration should be expressed as the sum of the independent aberrations from the uplink, downlink and the $\text{N}-1$ intermediate horizontal channels as  
\begin{equation}
    \phi(\Vec{r},t)=\phi_\text{u}(\Vec{r},t)+\sum_{i=1}^{N-1}\phi_i(\Vec{r},t)+\phi_\text{d}(\Vec{r},t).
\end{equation}
Thus the spatial variance of phase aberrations without AO system is given by
\begin{widetext}
\begin{equation}
    \sigma^2_{W_P}(\phi)=\sigma^2_\text{u}(\phi)+\sum_{i = 1}^{N - 1}\sigma^2_i(\phi)+\sigma^2_\text{d}(\phi)=\sum_{\mathrm{n},\mathrm{m}}\left[(b^{\mathrm{m}}_{\mathrm{n}})^2_\text{u}+\sum_{i = 1}^{N - 1}(b^{\mathrm{m}}_{\mathrm{n}})^2_i+(b^{\mathrm{m}}_{\mathrm{n}})^2_\text{d}\right].
\end{equation}
\end{widetext}
Here, there are two possible schemes for the AO system. One is the AO system only at the receiver of the downlink channel. The other is the AO system at the receiver of every channel section. In the former case, the phase variance can be written as
\begin{equation}
    \sigma^2_{W_P}(\phi)=\sum_{\mathrm{n},\mathrm{m}}(\gamma_{\mathrm{n}}^2)_\text{d}\left[(b^{\mathrm{m}}_{\mathrm{n}})^2_\text{u}+\sum_{i = 1}^{N - 1}(b^{\mathrm{m}}_{\mathrm{n}})^2_i+(b^{\mathrm{m}}_{\mathrm{n}})^2_\text{d}\right].
\end{equation}
In the latter case, the phase variance of the upper link is accumulated step by step in the subsequent links and corrected by the AO system, which is given by
\begin{widetext}

\begin{equation}
    \sigma^2_{W_P}(\phi)=\sum_{\mathrm{n},\mathrm{m}}\left[(\gamma_{\mathrm{n}}^2)_\text{u}(\gamma_{\mathrm{n}}^2)_\text{h}^{N - 1}(\gamma_{\mathrm{n}}^2)_\text{d}(b^{\mathrm{m}}_{\mathrm{n}})^2_\text{u}+\sum_{i = 1}^{N - 1}(\gamma_{\mathrm{n}}^2)_\text{h}^{N - i}(\gamma_{\mathrm{n}}^2)_\text{d}(b^{\mathrm{m}}_{\mathrm{n}})^2_i+(\gamma_{\mathrm{n}}^2)_\text{d}(b^{\mathrm{m}}_{\mathrm{n}})^2_\text{d}\right],
\end{equation}
\end{widetext}
where $(\gamma_{\mathrm{n}}^2)_\text{u}$, $(\gamma_{\mathrm{n}}^2)_\text{h}$, $(\gamma_{\mathrm{n}}^2)_\text{d}\in[0,1]$ denotes the mode attenuation factor in uplink, horizontal and downlink channels, respectively. Apparently, the latter case performs better on reducing the phase aberrations, which is adopted in our simulation and proposal.
According to Ref.~\cite{scriminich_optimal_2022}, the coefficients are independent and Gaussian-distributed as $b_{\mathrm{n}}^{\mathrm{m}}\sim\mathcal{N}\left(0,\langle b_{\mathrm{n}}^{\mathrm{m}2}\rangle\right)$, where $\langle b_{\mathrm{n}}^{\mathrm{m}}{}^2 \rangle$ is the Zernike coefficient variances given by Ref.~\cite{Dai:07}
\begin{equation}
\label{equalvariance}
\langle b_{\mathrm{n}}^{\mathrm{m}}{}^2 \rangle =
\frac{0.023 (\mathrm{n}+1) \Gamma(14/3) \Gamma(\mathrm{n} - 5/6) \pi^{8/3}}
{2^{5/3} [\Gamma(17/6)]^2 \Gamma(\mathrm{n} + 23/6)}
\left( \frac{D_{R_x}}{r_0} \right)^{5/3}.
\end{equation}
Here, $r_0$ is the Fried's parameter~\cite{andrews_laser_2005}. For uplink and downlink:
\begin{equation}
r_0 = 2.1\left[ \frac{\cos \theta_z}{1.46 k^2 \left( \mu_{1} + 0.622 \mu_{2} \Lambda^{11/6} \right)} \right]^{3/5},
\end{equation}
where
\begin{equation}
    \mu_1=\int_{h_0}^HC_n^2(h)\left[\Theta+\bar{\Theta}(1-\xi(h))\right]^{5/3}\diff h.
\end{equation}
For horizontal link:
\begin{equation}
r_0 = 2.1\left[ \frac{8}{3\left(a+0.62\Lambda^{11/6} \right)} \right]^{3/5}\left(1.46C^2_nk^2L_h\right)^{-3/5},
\end{equation}
where
\begin{equation}
a=\left\{
\begin{aligned}
\frac{1-\Theta^{8/3}}{1-\Theta},\ \ {\Theta \geq 0}\\ 
\frac{1+|\Theta|^{8/3}}{1-\Theta},\ \ {\Theta<0}
\end{aligned}
\right.
\end{equation}
Therefore, the average coupling efficiency related to phase aberrations corrected by the AO system is given by
\begin{widetext}

\begin{equation}
\label{equalAO}
    \langle \eta_{\text{AO}} \rangle =
    \prod_{\substack{\mathrm{n},\mathrm{m}}}\left[
    \frac{1}{\sqrt{1 + 2 (\gamma_{\mathrm{n}}^2)_\text{u}(\gamma_{\mathrm{n}}^2)_\text{h}^{N - 1}(\gamma_{\mathrm{n}}^2)_\text{d} \langle b_{\mathrm{n}}^{\mathrm{m}2} \rangle_\text{u}}}\cdot \prod_{\substack{i = 1}}^{N - 1}
    \frac{1}{\sqrt{1 + 2 (\gamma_{\mathrm{n}}^2)_\text{h}^{N - i}(\gamma_{\mathrm{n}}^2)_\text{d} \langle b_{\mathrm{n}}^{\mathrm{m}2} \rangle_i}} \cdot 
    \frac{1}{\sqrt{1 + 2 (\gamma_{\mathrm{n}}^2)_\text{d} \langle b_{\mathrm{n}}^{\mathrm{m}2} \rangle_\text{d}}} \right]. 
\end{equation}
\end{widetext}
According to the results shown in Fig.~\ref{efficiency}\textbf{(b)}, the implementation of AO further reduces channel losses by 25 dB at the distance of 10,000 km. 

\paragraph{$\eta_s$}
\label{Scintillation}
According to the result in Ref.~\cite{Canuet:18, Fried:66, Fried:67}, it can be written as
\begin{equation}
    \langle \eta_s \rangle \simeq \exp[-C_\chi(0)]=\exp[-\sigma^2_\chi],
\end{equation}
where $C_\chi(\rho)$ is the log-amplitude spatial covariance function and $C_\chi(0)=\sigma^2_\chi$ is the log-amplitude variance with 
\begin{equation}
    \exp[-\sigma^2_\chi]=\left(1+\sigma^2_I\right)^{-1/4},
\end{equation}
where $\sigma^2_I$ is the aperture-averaged scintillation index under spherical-wave approximation, given by Ref.~\cite{andrews_laser_2005} as
\begin{widetext}

\begin{equation} 
 \sigma^2_I=\exp\left[\frac{0.49\beta^2_0}{\left(1+0.18d^2+0.56\beta^{12/5}_0\right)^{7/6}}+\frac{0.51\beta^2_0\left(1+0.69\beta^{12/5}_0\right)^{-5/6}}{1+0.90d^2+0.62d^2\beta^{12/5}_0}\right]-1,
\end{equation} 
\end{widetext}
\begin{equation}
    \beta^{2}_0=0.4065\sigma^2_R,
\end{equation}
\begin{equation}
    d=\sqrt{\frac{kD^2_G}{4L_0}},
\end{equation}
where $D_G$ is the so-called hard aperture diameter of the receiver related to $D_{R_x}$ by $D^2_G=2D^2_{R_x}$.

Similar to phase aberrations, the overall log-amplitude fluctuations is the accumulation of the independent fluctuations from the uplink, downlink and the $\text{N}-1$ intermediate horizontal channels,
\begin{equation}
\chi(\vec{r},t)=\chi_\text{u}(\vec{r},t)+\sum_{i=1}^{\text{N}-1}\chi_i(\vec{r},t)+\chi_\text{d}(\vec{r},t).
\end{equation}
Therefore, the log-amplitude variance is given by
\begin{equation}
\sigma^2_\chi=(\sigma^2_\chi)_\text{u}+\sum_{i=1}^{\text{N}-1}(\sigma^2_\chi)_i+(\sigma^2_\chi)_\text{d},
\end{equation}
and the average coupling efficiency related to scintillation can be represented by
\begin{equation}
    \langle \eta_s \rangle=\left(1+\sigma^2_I\right)^{-1/4}_\text{u}\cdot \prod_{\substack{i=1}}^{\text{N}-1}\left(1+\sigma^2_I\right)^{-1/4}_i \cdot \left(1+\sigma^2_I\right)^{-1/4}_\text{d}.
\end{equation} 
%%%%%%%%%%%%%%%%%%%%%%%%%%%%%%%%%%%%%%%%%%%%%%%%%%%%%%%%%%%%%
Our simulations give an approximate efficiency $\eta_{\text{SMF}}$ of 72\% at 10,000 km, with slight variations depending on the free-space channel length.

\subsection{Parameter Exploration} 
\label{Parameter Exploration}
%%%%%%%%%%%%%%%%%%%%%%%%%%%%%%%%%%%%%%%%%%%%%%%%%%%%%%%%%%%
%%%%%%%%%%%% Extended Data Fig.3 %%%%%%%%%%%%%%%%%%%%%%%%%%%%%%%%%%
%%%%%%%%%%%%%%%%%%%%%%%%%%%%%%%%%
\begin{figure*}[t]
\centering
\includegraphics[width= 1\linewidth]{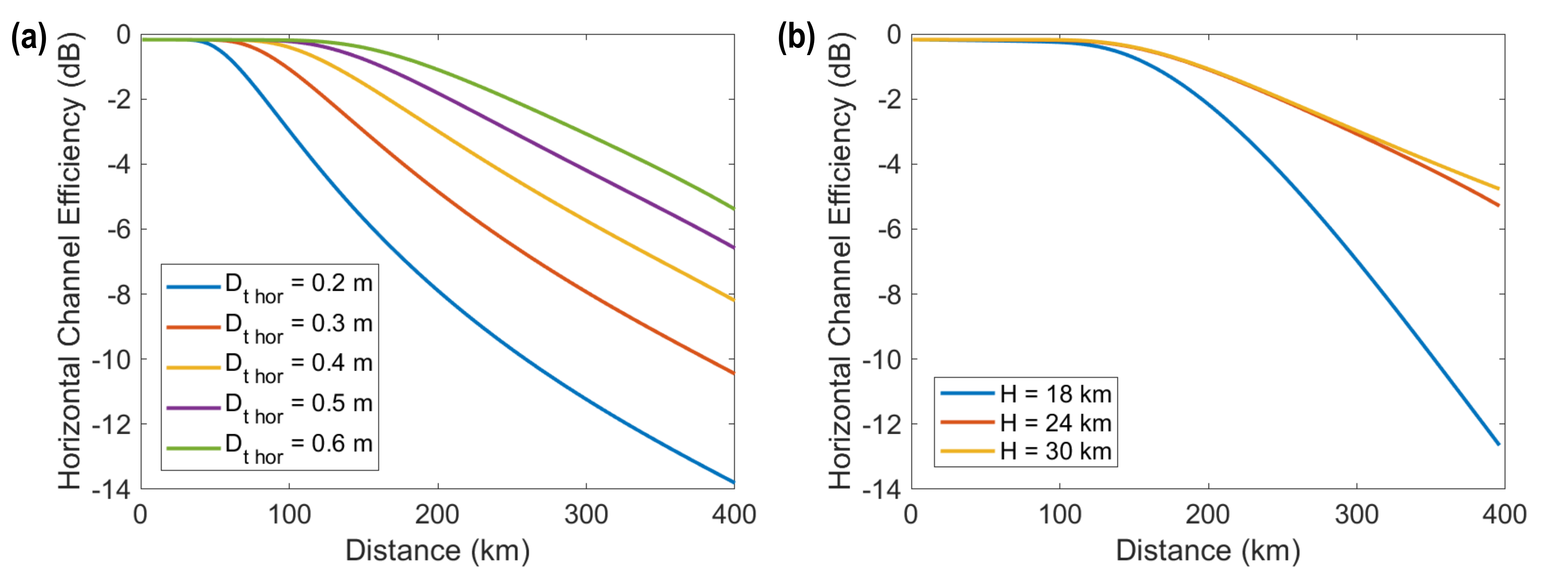}
\caption{Horizontal channel parameter walk with SMF coupling. 
\textbf{(a)} Channel efficiency versus $L_v$ with $H=24$ km, receiver diameter $D_\text{r hor}=0.6$ m, AO correction order $N_\text{AO}=10$.
\textbf{(b)} Channel efficiency versus $L_v$ with transmitted beam diameter $D_\text{t hor}=0.6$ m, receiver diameter $D_\text{r hor}=0.6$ m and $N_\text{AO}=10$. 
}
\label{figS3}
\end{figure*}
%%%%%%%%%%%%%%%%%%%%%%%%%%%%%%%%%%%%%%%%%%%%%%%%%%%%%%%%

%%%%%%%%%%%% Extended Data Fig.4 %%%%%%%%%%%%%%%%%%%%%%%%%%%%%%%%%%
\begin{figure*}[tb]
\centering
\includegraphics[width= 1.0 \linewidth]{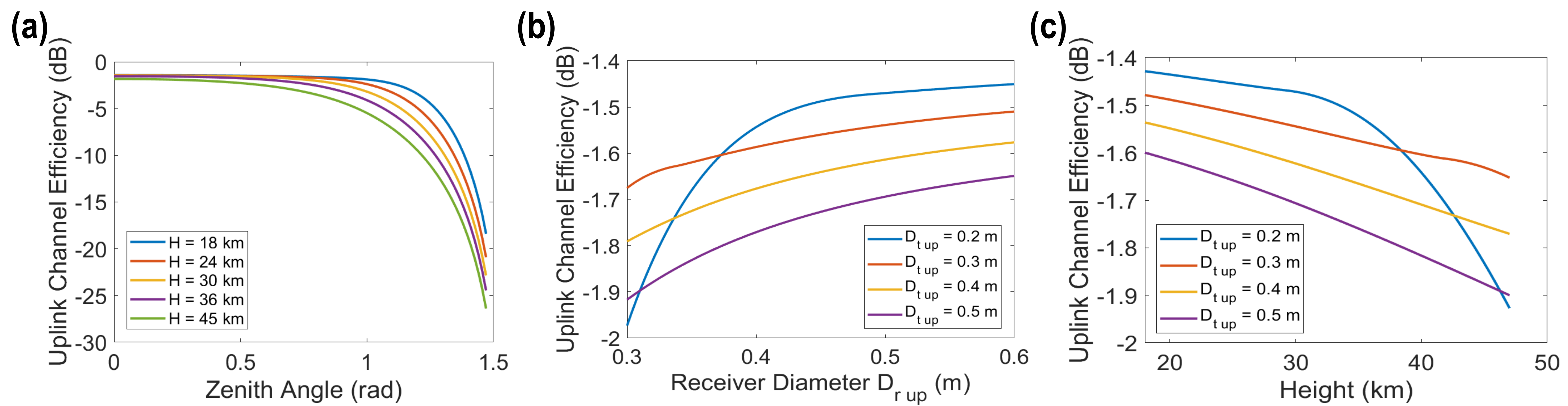}
\caption{Uplink parameter walk ($h_0=2$ m) with SMF coupling. \textbf{(a)} Channel efficiency versus zenith angles with $D_\text{t up}=0.2$ m, $D_\text{r up}=0.6$ m, $N_\text{AO}=10$, showing the optimal zenith angle $\theta_z=0$.  
\textbf{(b)} Channel efficiency versus $D_\text{r up}$ with $H=24$ km, $\theta_z=0$, $N_\text{AO}=10$. In practice, the size of telescopes on balloons should not be too large, therefore we limit $D_r$ to 0.6 m.
\textbf{(c)} Channel efficiency versus $H$ with $D_\text{r up}=0.6$ m, $\theta_z=0$, $N_\text{AO}=10$. 
}
\label{figS4}
\end{figure*}
%%%%%%%%%%%%%%%%%%%%%%%%%%%%%%%%%%%%%%%%%%%%%%%%%%%%%%%%

%%%%%%%%%%%% Extended Data Fig.5 %%%%%%%%%%%%%%%%%%%%%%%%%%%%%%%%%%
\begin{figure*}[tb]
\centering
\includegraphics[width= 1 \linewidth]{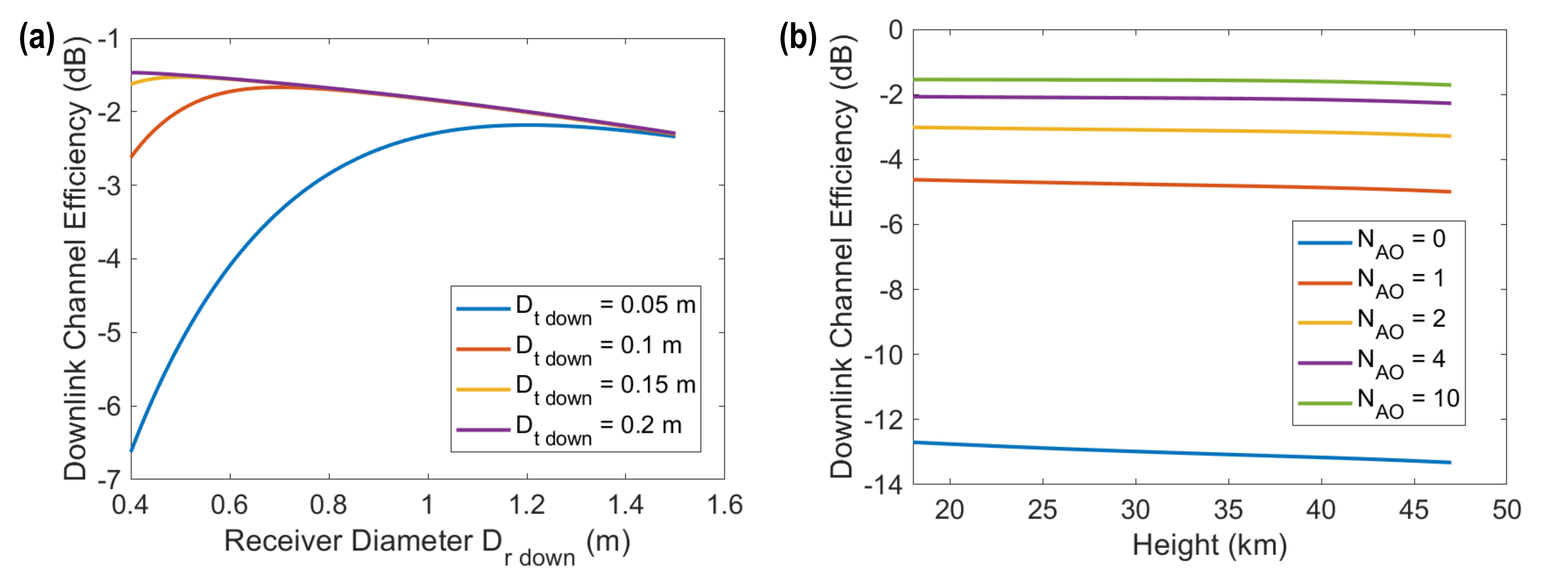}
\caption{Downlink parameter walk ($h_0=2$m) with SMF coupling. 
\textbf{(a)} Channel efficiency as a function of $D_\text{r down}$ , given $H=24$ km, $\theta_z=0$, $N_\text{AO}=10$. There is a trade-off between collection efficiency and coupling efficiency as discussed in Ref.~\cite{scriminich_optimal_2022}. Assume that the tracking system is not perfect, we take $D_\text{t down}=0.2$ m, $D_\text{r down}=0.6$ m as the preferable choice.
\textbf{(b)} Channel efficiency as a function of $H$ with varying $N_\text{AO}$, given $D_{t down}=0.2$ m, $D_\text{r down}=0.6$ m, $\theta_z=0$, demonstrating that the AO system significantly enhances channel efficiency.
} 
\label{figS5}
\end{figure*}

%%%%%%%%%%%%%%%%%%%%TABLE 1%%%%%%%%%%%%%%%%%
%%%%%%%%%%%%%%%%%%%%TABLE 1%%%%%%%%%%%%%%%%%
\begin{table*}
    \centering
    
    \textbf{Channel Parameters}\\

    \begin{ruledtabular}
    \begin{tabular}{ccc}
        Parameter & 
        Value & 
        Description \\ 
        \hline
        
        $\lambda$ & 
        $1537$ nm & 
        Wavelength \\
        
        $\eta_t, \eta_r$ & 
        $0.98$ & 
        Optical transmittance of transmitter, receiver \\
        
        $D_t$ (uplink/downlink channel) & 
        $0.2$ m & 
        Uplink/downlink transmitted beam diameter \\

        $D_t$ (horizontal channel) & 
        $0.6$ m & 
        Horizontal channel transmitted beam diameter \\

        $D_r$ (uplink/downlink/horizontal channel) & 
        $0.6$ m & 
        Uplink/downlink/horizontal channel receiver diameter \\
        
        $H$ & 
        $24$ km & 
        Height of balloons \\
        
        $h_0$ & 
        $2$ m & 
        Altitude of ground stations \\
        
        $\theta_z$ &
        $0$ &
        Zenith angle \\
    \end{tabular}
    \end{ruledtabular}

    \vspace{1em} 

    \textbf{AO System Parameters (similar as that in Ref. \cite{PhysRevResearch.7.023199} except $N_\text{AO}$)} 
    \begin{ruledtabular}
    \begin{tabular}{ccc}
        Parameter & Value & Description \\ 
        \hline
        $N_\text{AO}$ & $10$ & Maximum radial order corrected by AO system \\
        
        $\bar{v}_\text{wind}$ &
        $10$ m/s &
        Average transverse wind velocity \\
        
        $K_I$ & 
        1 &
        Integral gain of the AO system \\
        
        $\tau_{AO}$ &
        $2.10^{-3}$ s &
        Control delay of the AO system \\
        
        $T_{AO}$ &
        $1.10^{-3}$ s &
        Integration time of the AO system \\
    \end{tabular}
    \end{ruledtabular}
 
    \caption{\label{tab:simulation_parameters} Simulation Parameters}
    
\end{table*}
%%%%%%%%%%%%%%%%%%%%%%%%%%%%%%%%%%%%%%%%%%%%%%%%%%%%%%%%%%%%%%
To identify the optimal parameters for minimizing channel loss in our simulation, we investigated various factors, including different zenith angles, balloon altitudes, transmitter and receiver aperture sizes, and the correction orders of the adaptive optics (AO) system, as illustrated in Figs. \ref{figS3}–\ref{figS5}. Specifically, $D_t, D_r$ refer to $2W_0$ \text{(transmitted beam diameter)}, $D_{R_x} $\text{(receiver diameter)} discussed above, respectively. 

In the horizontal channel (Fig.~\ref{figS3}), the larger transmitted beam size and the higher balloons lead to better performances in channel efficiency. We consider the diameter up to 0.6 m for easy deployment of telescopes on balloons. Moreover, the efficiency improvement brought by the balloon's altitude is not significant when higher than 24 km since the impact of atmosphere attenuation and turbulence becomes negligible.

In the uplink channel (Fig.~\ref{figS4}), the efficiency decreases as the zenith angle and the height of balloons grow while increases versus the receiver diameter. The results indicate that the balloons better stay vertically above the ground observatories. Also, transmitted beam size, receiver diameter and the height of balloons should be comprehensively considered to balance the effect of beam diffraction, broadening and wandering as well as the wavefront distortion. Combining the results discussed above, we set the height of balloons to be 24 km.

In the downlink channel (Fig.~\ref{figS5}), we show the trade-off between collection efficiency and coupling efficiency (Ref.~\cite{scriminich_optimal_2022}) since larger receiver aperture improves the $\eta_{D_{R_x}}$ (Eq.~\ref{equalcollection}) but suppresses the $\eta_\text{AO}$ by increasing the aberration variance $\langle b_{\mathrm{n}}^{\mathrm{m}}{}^2 \rangle$ (Eq.~\ref{equalvariance}, ~\ref{equalAO}). Thus we should take the optimal $D_r$ to reach the balance. The channel efficiency is improved by at least 10 dB with the AO system, emphasizing the necessity of wavefront corrections.

According to these results, the channel parameters used in our simulation are listed in Table.~\ref{tab:simulation_parameters}.
%%%%%%%%%%%%%%%%%%%%%%%%%%%%%%%%%%%%%%%%%%%%%%%%%%%%%%%%%%%%%%%%%%%%%%%%%%%%%
 
%%%%%%%%%%%%%%%%%%%%%%%%%%%%%%%%%%%%%%%%%%%%%%%%%%%%%%%%%%%%%%%%%%%%%%%%%%%
\subsection{Discussion on Position Jitter of the Balloons} 
\label{Unideal case of balloons}
In the above simulation, we assume that all relay balloons are evenly distributed along the Earth's great-circle. However, as discussed in ``Technical Feasibility" section, for practical applications, we must also consider the dynamic position jitter of the balloons. 
Based on the results in Ref.~\cite{XU2022733}, reinforcement learning enables balloons to maintain station within approximately 10 km horizontally and 1 km vertically. To account for the resulting position jitter, we adopt a simplified two-dimensional random-walk model, assuming a fixed altitude of 24 km.

%To model this effect, we employ a simplified two-dimensional random walk, assuming that the balloons maintain a stable altitude of 24 km.
 
\begin{figure*}[tb]
\centering
\includegraphics[width= 1.0 \linewidth]{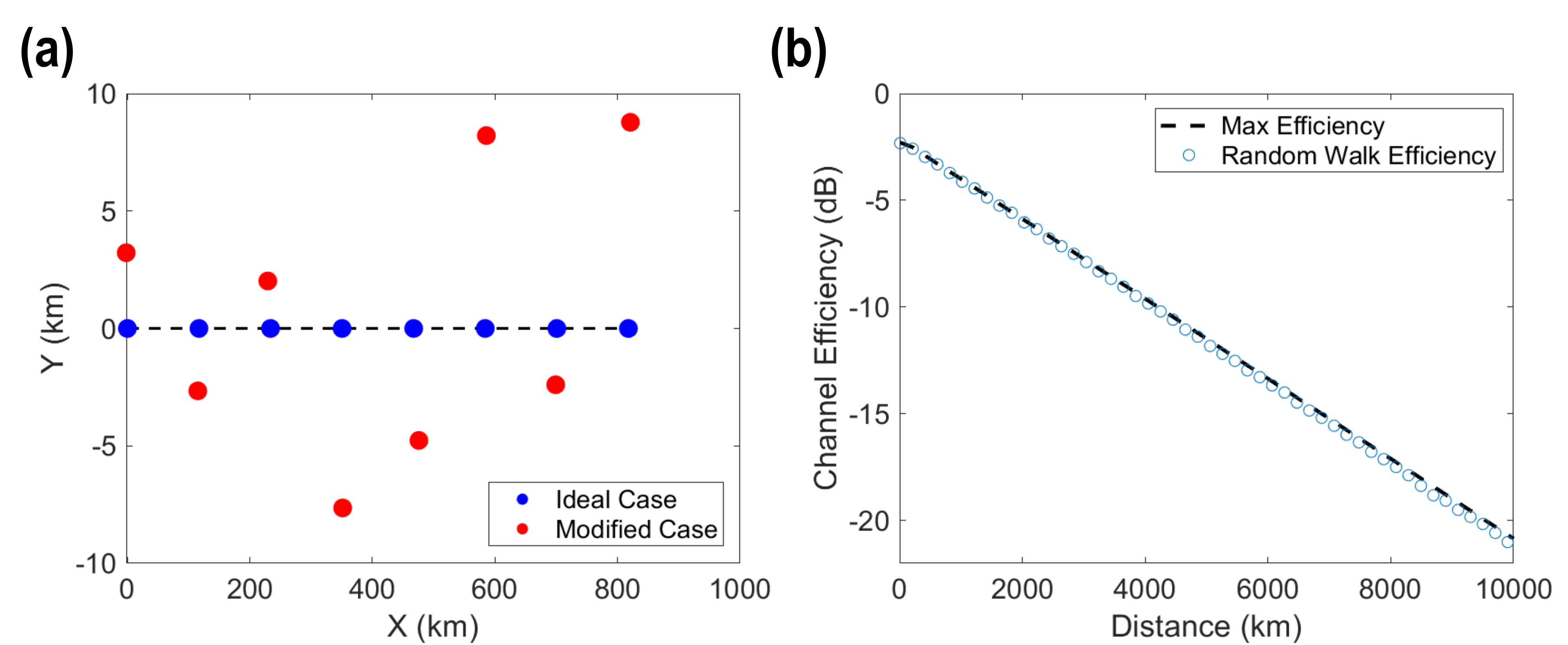}
\caption{\textbf{(a)} Modified model and \textbf{(b)} modified efficiency by random walking on balloons' positions with the optimal relay number. The ideal case is also shown for comparison. The position deviations in both \textbf{(a)} and \textbf{(b)} follow $(\Delta{x_i})^2+(\Delta{y_i})^2\leq10$ km, i=1,2,…,N.
}
\label{figS6}
\end{figure*}

An example of our model is depicted in Fig. \ref{figS6}\textbf{(a)}. Ideally, the optimal number (N) of balloons at a certain distance (L) are evenly spaced along the $y=0$ axis, with adjacent balloons separated by L/(N-1). Thus, the position of the i-th balloon (i=1,2,…,N) can be represented by ($x_i, 0$). The position jitter can be modeled by adding uniformly distributed random numbers as 
\begin{equation}
    (x'_i, y'_i) = (x_i + \Delta{x_i}, \Delta{y_i}), i=1,2,…,N
\end{equation}
where $\Delta{x_i}$ and $\Delta{y_i}$ are uniformly distributed within the range $(\Delta{x_i})^2+(\Delta{y_i})^2\leq10$ km. 
%This will lead to non-zero zenith angles, reducing the uplink and downlink channel efficiency. For the other relay balloons that influence the horizontal channel efficiency, both the uneven distribution of observatories and position fluctuations must be taken into account. Specifically, the position of the i-th relay balloon is adjusted as follows:
%where we assume the random numbers $\Delta{x_i}$ and $\Delta{y_i}$ are uniformly distributed within the range $\Delta{x_i},\Delta{y_i}\in[-0.2L/(N-1), 0.2L/(N-1)]$ in Fig.~\ref{figS6}\textbf{(b)} and $\Delta{x_i},\Delta{y_i}\in[-0.4L/(N-1), 0.4L/(N-1)]$ in Fig.~\ref{figS6}\textbf{(c)}, supposing the positional deviation increases as the distance grows.
Subsequently, we compute the modified zenith angles for the uplink and downlink channel efficiency, as well as the modified distances between neighboring balloons on the 2D plane for the horizontal channel efficiency. The results are depicted in Fig. \ref{figS6}\textbf{(b)}, revealing less than 0.5 dB decrease in channel efficiencies at 10,000 km compared to the ideal scenario.

%%%%%%%%%%%% Extended Data Fig.6 %%%%%%%%%%%%%%%%%%%%%%%%%%%%%%%%%%
%%%%%%%%%%%%%%%%%%%%%%%%%%%%%%%%%%%%%%%%%%%%%%%%%%%%%%%%

\begin{figure*}[tb]
\centering
\includegraphics[width= 0.45 \linewidth]{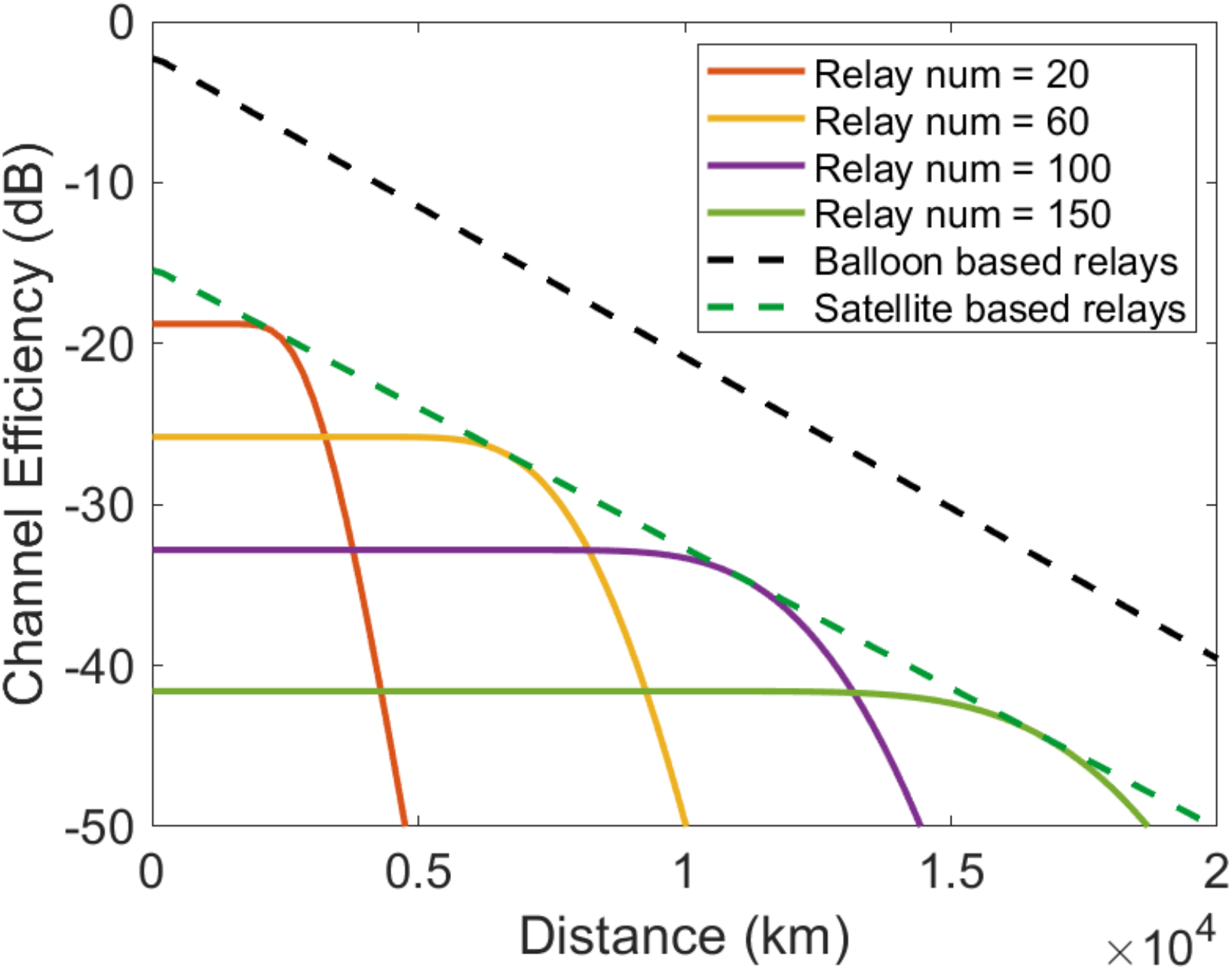}
\caption{ 
Channel efficiency of satellite-based relays as a function of the distance between servers, based on parameters provided in Table~\ref{tab:simulation_parameters} and Table~\ref{tab:satellite_parameters}. Solid lines with different colors represents various number of relay satellites. The green dashed line indicates maximum channel efficiency across all possible relay numbers, with the optimized channel efficiency of balloon-based relays (black dashed line) as a reference.
}
\label{figS7}
\end{figure*}
%%%%%%%%%%%%%%%%%%%%%%%%%%%%%%%%%%%%%%%%%%%%%%%%%%%%%%%%

%%%%%%%%%%%%%%%%%%%%TABLE 1%%%%%%%%%%%%%%%%%
\begin{table*}
    \centering
    
    \textbf{Channel Parameters}\\

    \begin{ruledtabular}
    \begin{tabular}{ccc}
        Parameter & 
        Value & 
        Description \\ 
        \hline
        
        $D_t$ (uplink channel) & 
        $1.2$ m & 
        Uplink transmitted beam diameter \\

        $D_r$ (uplink channel) & 
        $0.6$ m & 
        Uplink receiver diameter \\

        $D_t$ (downlink channel) & 
        $0.6$ m & 
        Downlink transmitted beam diameter \\

        $D_r$ (downlink channel) & 
        $1.2$ m & 
        Downlink receiver diameter \\
        
        $H$ & 
        $500$ km & 
        Height of satellites \\
        
    \end{tabular}
    \end{ruledtabular}

    \caption{\label{tab:satellite_parameters} Adjusted Simulation Parameters on Satellite Relays}
    
\end{table*}
%%%%%%%%%%%%%%%%%%%%%%%%%%%%%%%%%%%%%%%%%%%%%%%%%%%%%%%%%%%

%%%%%%%%%%%%%%%%%%%%%%%%%%%%%%%%%%%%%%%%%%%%%%%%%%%%%%%%%%%%%
\subsection{Performance of Satellite-based Relays}
\label{Satellite}
Satellite-based relays represent another potential free-space channel for H4QR. Consequently, we have conducted a channel efficiency simulation for satellite-based relays, with the results presented in Fig.~\ref{efficiency}\textbf{(a)} and Fig.~\ref{figS7}. It is important to note that our simulation is more rigorous than the one reported in Ref.~\cite{goswami_satellite-relayed_2023}, as we have additionally considered wavefront distortion in both uplink and downlink channels. We have also applied the optimization strategy of beam waist positioning and AO correction, which makes our results more robust.

%%%%%%%%%%%% Extended Data Fig.7 %%%%%%%%%%%%%%%%%%%%%%%%%%%%%%%%%%

The overall calculation idea is similar to that of the balloons, with a few differences. First, the height of the satellites is set at 500 km, differing from the balloons. Thus, the effects of the atmosphere, such as attenuation, turbulence, and scintillation, are ignored in horizontal channels, leaving only optical and diffraction loss. Second, the pointing error resulting from the random drift of balloons is not considered for satellites, as their positions are substantially more stable and predictable. Third, we reevaluate the sizes of the transmitted beam and receiver apertures for the uplink and downlink channels to achieve higher efficiency. The adjusted parameters are shown in Table~\ref{tab:satellite_parameters}; those not listed remain the same as in Table~\ref{tab:simulation_parameters}.

Comparing the two dashed lines in Fig.~\ref{figS7}, the slope of satellite-based relays is slightly smaller than that of balloon-based relays due to the absence of atmosphere in the horizontal channels for satellites. However, the much higher altitude of satellites results in greater channel losses in the uplink and downlink channels, leading to a much lower intercept for satellite-based relays. The combination of these effects results in better channel efficiency for balloon-based relays, even for a pure relay global-coverage distance of 20,000 km. Consequently, balloon-based relays are more favorable for H4QR and global-scale networking.
%%%%%%%%%%%%%%%%%%%%%%%%%%%%%%%%%%%%%%%%%%%%%%%%%%%%%%%%%%%%%%%%%%

%%%%%%%%%%%%%%%%%%%%%%%%%%%%%%%%%%%%%%%%%%%%%%%%%%%%%%%%%%%%
\section{Calculations of Distribution Time of H4QR.}
\label{section:Calculations of Distribution time of H4QR}
We start with a simplified two-segment model for H4QR, assuming that the EPPS is sufficient to fill the QM capacity, and the entanglement generation between clients and local servers is much faster than the entanglement generation between servers, so that the entanglement stored in QM2/QM5 is always ready. Such simplification is adequate only for long distances (see definition in the next paragraph). For shorter distances, we introduce corrections to this simplified model by further considering the restriction of repetition rates of EPPS and the channel loss of fiber links.
%Further in Appendix, we give a more accurate analysis for short-distance distribution by adjusting model parameters, considering the restriction of EPPS repetition rate and fiber channel loss.
%Let's start with a simplified model for long-distance distribution, assuming entanglement generation between clients and servers is significantly faster than that through free-space links, therefore QM 2 and 5 are always prepared for operating BSMs. 

% This underscores the critical role of multiplexing in both the frequency and spatial domains~\cite{yang_multiplexed_2018,ou2025multichannelhighdimensionalintegrated,PhysRevLett.113.053603}.

The EPPS generates entangled pairs with a repetition rate $R$ and probability $\rho$. The system runs with a clock $L/c$, where $L$ is the distribution distance between Charlie and David, and $c$ is the light speed in the communication channel. For long distances scenarios satisfying $L>cnm/R$, the EPPS generates sufficient entangled photon-pairs in one system clock to fully load the QM capacity ($RL/c>nm$, where $n$ and $m$ are independent and dependent mode number of QMs, as defined in the main text). Under this condition, the entanglement distribution rate becomes independent of the photon source repetition rate $R$. Notably, at such extended distances, the transmission efficiency of free-space channels is significantly lower than that of metropolitan fiber channels, further justifying our simplified two-segment model.  
      
Based on the above-mentioned simplifications, the H4QR architecture can be modeled as a multiplexed quantum repeater comprising only two segments (Charlie $\leftrightarrow$ Elbert, denoted as A, and David $\leftrightarrow$ Elbert, denoted as B). For such systems, O. A. Collins et al. \cite{collins_multiplexed_2007} have derived an approximate solution for EDR as,

\begin{equation}
\begin{aligned}
f_{\tau, n} & =\frac{P_1\left(1-q_0^n\right)\left(1+q_0^n-2 q_0^{n(\tau+1)}\right)}{1+2 q_0^n-q_0^{2 n}-4 q_0^{n(\tau+1)}+2 q_0^{n(\tau+2)}+\alpha}, \\
\alpha & =\frac{q_0^{n-1}\left(1-q_0^n\right)\left[1-q_0^{2 n-1}+2 q_0^{3 n-2}\left(1-q_0^{\tau(2 n-1)}\right)\right]}{\left(1-q_0^{2 n-1}\right)\left(1+q_0^n-2 q_0^{(\tau+1) n}\right)} ,
\end{aligned}
\end{equation}  
under the condition $np\ll1$, where $p$ represents the entanglement generation probability in a single independent mode, and $q_0=1-p$. In the specific case of H4QR with $m$ dependent modes, $p\equiv p_m$ is given by:
\begin{equation} 
p_m=1-\left[1-\eta_\text{ch}\eta_\text{M}\rho^2\left(\frac{1}{2}\eta_\text{D}^2\right)^2\right]^m
\end{equation}
with $\eta_\text{ch}$ the free-space channel efficiency per segment while $\eta_\text{M}$ and $\eta_\text{D}$ represent the efficiencies of the QM and SPD, respectively. The distribution time is subsequently derived as $T=\frac{L}{c}f_{\tau, n}^{-1}$.
It should be noted that this approximation assumes no simultaneous successful entanglement generation events across different modes—an assumption that becomes invalid as the mode number $n$ increases. Nevertheless, it provides an upper bound for the entanglement distribution time.
 
We further propose another method to derive a lower bound. We assume that the lifetime of the quantum memory $\tau$ is much longer than the system clock $L/c$, which is consistent with practical conditions where $\tau$ is approximately $1$ s. Define a random variable $\varepsilon$ as the number of residual entanglements after one system clock. The residual entanglements can only exist in one of the segments, which means $\varepsilon\in [0,n]$, since entanglement swapping at BSM4 is carried out when entanglements are established in both segments for at least one mode. Without loss of generality, we assume the residual entanglement exists in segment A. With long-lifetime quantum memories in which residual entanglements won't perish spontaneously, $\varepsilon=\varepsilon+1$ when entanglement is created in segment A, and $\varepsilon=\varepsilon-1$ only when entanglement for one of these modes is created in segment B. Therefore, this problem is transformed into a one-dimensional random walk along the number axis of $\varepsilon$. The transition matrix can be derived as follows,
    
\begin{equation}
	\mathcal{T}_{ij}=\left\{
	\begin{aligned}
		&(1-p_m)^{2n-j+1}  & i=j\\
		&\frac{n}{2n-j+1}(1-(1-p_m)^{2n-j+1})   & i=j-1\\
            &\frac{n-j+1}{2n-j+1}(1-(1-p_m)^{2n-j+1})   & i=j+1 \wedge j\neq 1\\
            &1-(1-p_m)^{2n} & i=2,j=1\\
            &0 &\mathrm{otherwise}
	\end{aligned}
	\right.
 \label{eq.trans matrix}
\end{equation} 
where $\mathcal{T}_{ij}$ represents the transfer probability from $\varepsilon=j-1$ to $\varepsilon=i-1$, and $i,j\in [1,n+1]$.
Then the probability distribution $P_n(\varepsilon)$ can be derived by solving for the normalized eigenvector with eigenvalue of 1. Noting that the whole system runs with $2n-\varepsilon$ modes, the expected number of operating modes is $\sum^{n}_{\varepsilon=0}(2n-\varepsilon)P_n(\varepsilon)$, and the distribution time is thus given by
\begin{equation}
	\langle T \rangle = \frac{2L}{cp_mP_{1}}\left[\sum^{n}_{\varepsilon=0}(2n-\varepsilon)P_n(\varepsilon)\right]^{-1},
\label{Eq.T}
\end{equation}
where the factor $2$ appears as two successful entanglement generation leads to one attempt of entanglement swapping, and $P_1=\frac{1}{2}\eta_\text{M}^2\eta_\text{D}^2$ represents the success probability of entanglement swapping at BSM4.

%It requires the modes to be stored and retrieved independently for multiplexing. In practical QMs, the temporal modes can only be stored and retrieved together. As a result, the temporal modes number $m$ only improve the success probability of every attempt of entanglement generation as $p_m=1-(1-p)^m$, with $p=\eta_{link}\eta_\text{M}\rho^2(\frac{1}{2}\eta_\text{D}^2)^2$ in HCQRBR, where $\eta_{link}$ denotes efficiency of free-space link of a segment, and $\rho$ represents the photon-pair generation probability of EPPS.
In the analysis above, the repetition rate of entanglement generation is assumed to be limited only by the capacity of QMs. 
%However in the case of short distances, i.e. $RL/c<nm$, the EPPS can not fully utilize the memory capacity. Furthermore, at shorter distances, the efficiency of free-space channels becomes comparable to that of metropolitan fiber channels, rendering the two-segment assumption no longer valid. To make more reliable predictions of EDR for H4QR at short distances, we modify this simplified model by replacing $p_m$ with a function $p(m,k,L)$, and correspondingly adjust the form of $\mathcal{T}_{ij}$ and $\langle T \rangle$, with details provided in the Appendix. Moreover, a Monte Carlo simulation is performed to compare with the theoretical results.
However, in the case of short distances where $RL/c<nm$, the entangled photon pairs generated per clock cycle cannot saturate the QM capacity. To derive the upper bound of distribution time, we run the system slower with a clock of $\max\{\frac{nm}{R},\frac{L}{c}\}$, which leads to $T=\max\{\frac{nm}{R},\frac{L}{c}\}f_{\tau, n}^{-1}$. This configuration ensures in the distance of $L<\frac{cnm}{R}$, the EPPS operates within its full capacity with a fixed system clock. 

The diminishing efficiency gap between free-space and fiber channels at short distances necessitates careful consideration of fiber channel loss. Given that $L_f\ll \frac{cnm}{R}$, QM capacity constraints become negligible for client-server entanglement generation. Noting the entanglement generation rate between clients and servers follows $R_f=\frac{1}{2}\eta_\text{D}^2e^{-\alpha L_f}\rho^2R$, we only need to restrict the entanglement generation rate between immediate servers and clients to not exceeding $\frac{1}{2}\eta_\text{D}^2\eta_\text{M}R_f$, where the factor $\frac{1}{2}\eta_\text{D}^2\eta_\text{M}$ represents the success probability of BSM2/BSM6.
To account for this constraint, we can equivalently modify the expression of $p_m$ as follows,
\begin{equation}
    \begin{aligned}
    p_{m}=\min\Bigg\{1-\left[1-\eta_\text{ch}\eta_\text{M}\rho^2\left(\frac{1} {2}\eta_\text{D}^2\right)^2\right]^{m},
    \\ \frac{1}{2}\eta_\text{D}^2\eta_\text{M}\left[1-\left(1-\frac{1}{2}\eta_\text{D}^2\rho^2e^{-\alpha L_f}\right)^m\right] \Bigg\}.
    \end{aligned}
\end{equation}
\\

To determine the lower bound of short distance distribution time, we maintain the original system clock $\frac{L}{c}$, while introducing adaptive mode allocation as:
\begin{equation}
m_\varepsilon=\min\{\frac{RL}{c(n-\varepsilon)} ,m\}.
\end{equation}

This optimization of EPPS utilization allocates entangled photon-pairs equally to each independent mode, with $\varepsilon$ modes occupied by the residual entanglements. Then the $p_m$ of the segment with residual entanglement (segment A) in Eq.~\eqref{eq.trans matrix} is revised as follows,
\begin{widetext}
\begin{equation}
\begin{aligned}
p_{m,\varepsilon}=\min\Bigg\{1-\left[1-\eta_\text{ch}\eta_\text{M}\rho^2\left(\frac{1} {2}\eta_\text{D}^2\right)^2\right]^{m_\varepsilon}, \frac{1}{2}\eta_\text{D}^2\eta_\text{M}\left[1-\left(1-\frac{1}{2}\eta_\text{D}^2\rho^2e^{-\alpha L_f}\right)^{m_\varepsilon}\right]\Bigg\},
\end{aligned}
\end{equation}
\end{widetext}
and in segment B there is no residual entanglement, therefore $p_{m,\varepsilon}=p_{m,0}$ for that segment. Noting that $\varepsilon=j-1$ in matrix element $\mathcal{T}_{ij}$, we can derive the corrected matrix element as:
\begin{widetext}
\begin{equation}
	\mathcal{T}_{ij}=\left\{
	\begin{aligned}
		&(1-p_{m,0})^n(1-p_{m,j-1})^{(n-j+1)}  & i=j\\
		&\frac{np_{m,0}}{np_{m,0}+(n-j+1)p_{m,j-1}}(1-(1-p_{m,0})^n(1-p_{m,j-1})^{(n-j+1)})   & i=j-1\\
            &\frac{(n-j+1)p_{m,j-1}}{np_{m,0}+(n-j+1)p_{m,j-1}}(1-(1-p_{m,0})^n(1-p_{m,j-1})^{(n-j+1)})   & i=j+1 \wedge j\neq 1\\
            &1-(1-p_{m,0})^{2n} & i=2,j=1\\
            &0 &\mathrm{otherwise}
	\end{aligned}
	\right.
 \label{eq.trans matrix2}
\end{equation}
\end{widetext}

This formulation enables precise calculation of distribution time:
\begin{equation}
	\langle T \rangle = \frac{2L}{cP_{1}}\left[np_{m,0}+\sum^{n}_{\varepsilon=0}(n-\varepsilon)p_{m,\varepsilon}P_n(\varepsilon)\right]^{-1}
\end{equation}
which demonstrates excellent agreement with Monte Carlo simulation result as shown in Fig.~\ref{edr}. And one should note that millions of experiments are needed in Monte Carlo simulation in the case of small $m$ to obtain a stable result.
%%%%%%%%%%%%%%%%%%%%%%%%%%%%%%%

%%%%%%%%%%%%%%%%%%%%%%%%%%%%%%%%%%%
\section{Detailed Analysis of Entanglement Distribution Time}
\label{SMsec2}
%%%%%%%%%%%%%%%%%%%%%%%%%%%%%%%%%%%%%%%%%%%%%%%%%%%%%
\subsection{Detailed Analysis of Parameters Sensitivity}
\label{section EDR vs paras}

\begin{figure*}[t]
\centering
\includegraphics[width= 1 \linewidth]{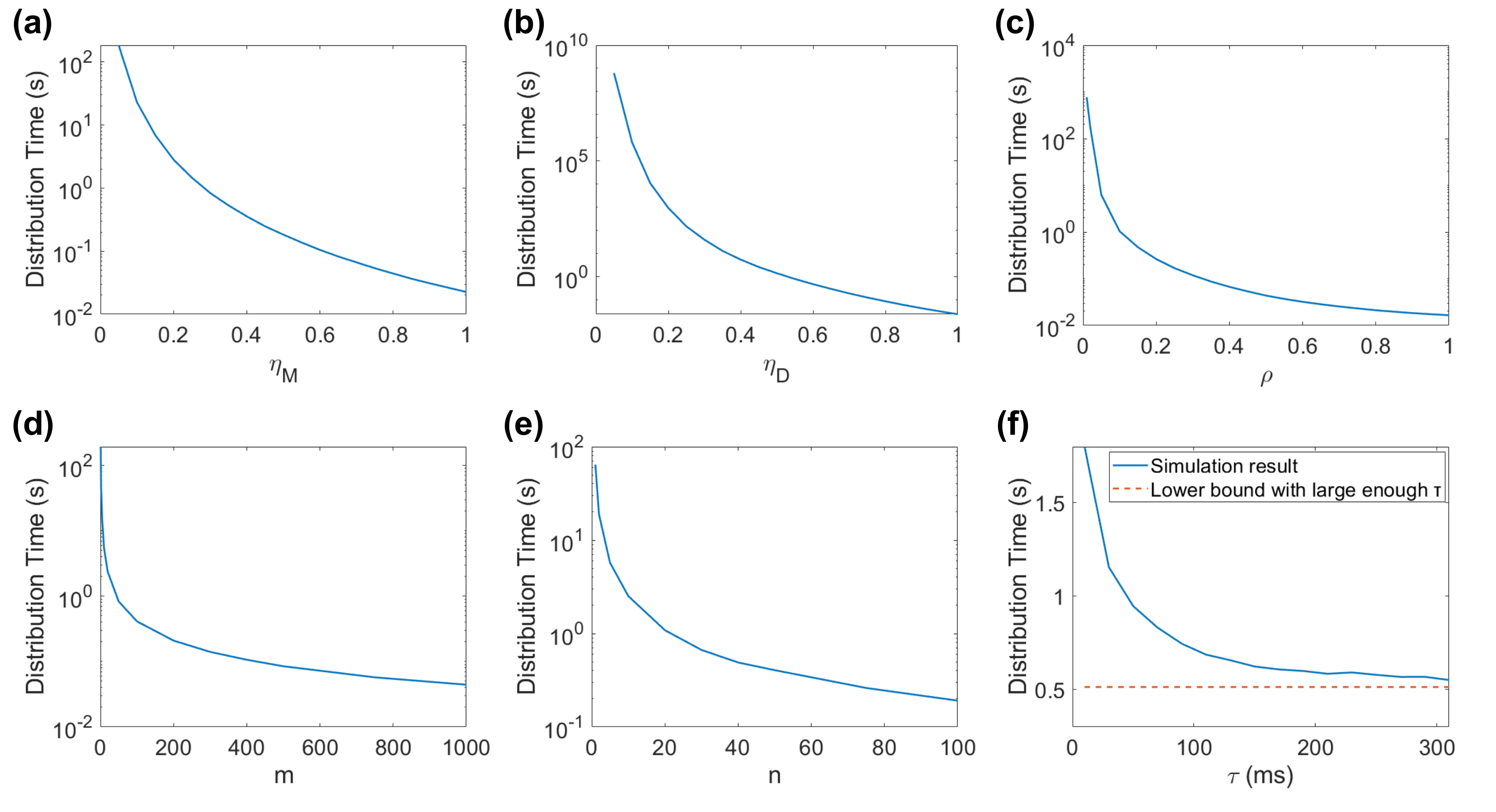}
\caption{Distribution time of H4QR varying with parameters $\eta_\text{M},\eta_\text{D},\rho,m,n,\tau$. All figures draws with 3000 km distribution distance, 25 km metropolitan fiber channels and 1 MHz EPPS repetition rate.
\textbf{(a)} Distribution time as a function of $\eta_\text{M}$, with $\eta_\text{D}=0.9,~\rho=0.05,~m=1000,~n=10,$ and$~\tau=1$ s.
\textbf{(b)} Distribution time as a function of $\eta_\text{D}$, with $\eta_\text{M}=0.8,~\rho=0.05,~m=1000,~n=10,$ and$~\tau=1$ s.
\textbf{(c)} Distribution time as a function of $\rho$, with $\eta_\text{M}=0.8,~\eta_\text{D}=0.9,~m=10,~n=10,$ and$~\tau=1$ s.
\textbf{(d)} Distribution time as a function of $m$, with $\eta_\text{M}=0.8,~\eta_\text{D}=0.9,~\rho=0.05,~n=10,$ and$~\tau=1$ s.
\textbf{(e)} Distribution time as a function of $n$, with $\eta_\text{M}=0.8,~\eta_\text{D}=0.9,~\rho=0.05,~m=20,$ and$~\tau=1$ s.
\textbf{(f)} Distribution time as a function of $\tau$, with $\eta_\text{M}=0.8,~\eta_\text{D}=0.9,~\rho=0.05,~n=10,$ and$~m=100$.
}
\label{fig:EDR vs pars}
%label要放在最下面，否则hyperref包会报错！
\end{figure*}

To assess the near-term feasibility, we conducted Monte Carlo simulations to analyze the performance of H4QR over a distance of 3000 km—a range that already exceeds the operational range of a single low-Earth-orbit (LEO) satellite—while varying key parameters.
    
%Conventional quantum repeaters suffer from a serious decline of EDR $(\frac{1}{2}\eta_\text{M}^2\eta_\text{D}^2)^N$ through N entanglement swapping operations. 
While conventional quantum repeaters necessitate a large number of quantum memories (QMs), H4QR significantly reduces this requirement, thereby achieving robust tolerance to $\eta_\text{M}$. As shown in Fig.~\ref{fig:EDR vs pars} (a) and (b), a sub-second distribution time remains attainable even when $\eta_\text{M}=0.3$.

Photon pair generation probability $\rho$ and dependent modes capacity of QM $m$ jointly determine the success probability of entanglement generation $p_m$. While a large enough $m$ enables nearly deterministic entanglement generation, a quasi-deterministic EPPS could be a more promising choice given practical limitations on repetition rate $R$. Note that Fig.~\ref{fig:EDR vs pars} (c) is calculated with only $m=10$, yet a quasi-deterministic EPPS outperforms hundreds of dependent QM modes.

%It is challenging to combine all state-of-the-art QM performance in a single system. Cavities enhance QM efficiency but restrict memory bandwidth~\cite{sabooni_cavity-enhanced_2013}. 

In our recent work~\cite{meng2025efficientintegratedquantummemory}, 20 temporal modes were stored in a cavity-enhanced QM with a bandwidth of 6 MHz. Spatial multiplexing can coexist with cavities using waveguide structures. As shown in Fig.~\ref{fig:EDR vs pars} (e), H4QR maintains a distribution time less than 2.5 s even with $n=10$ and $m=20$.

Multiplexed quantum repeaters are insensitive to QM lifetimes, as already discussed in Ref.~\cite{collins_multiplexed_2007}. Fig.~\ref{fig:EDR vs pars} (f) reveals that for 3000 km entanglement distribution, a memory time of 50 ms only doubling the distribution time compared to ideal infinite-lifetime cases.

\subsection{Multiplexing with Dependent and Independent Modes}
\label{section mn multip}
 
\begin{figure*}[tb]
\centering
\includegraphics[width= 1 \linewidth]{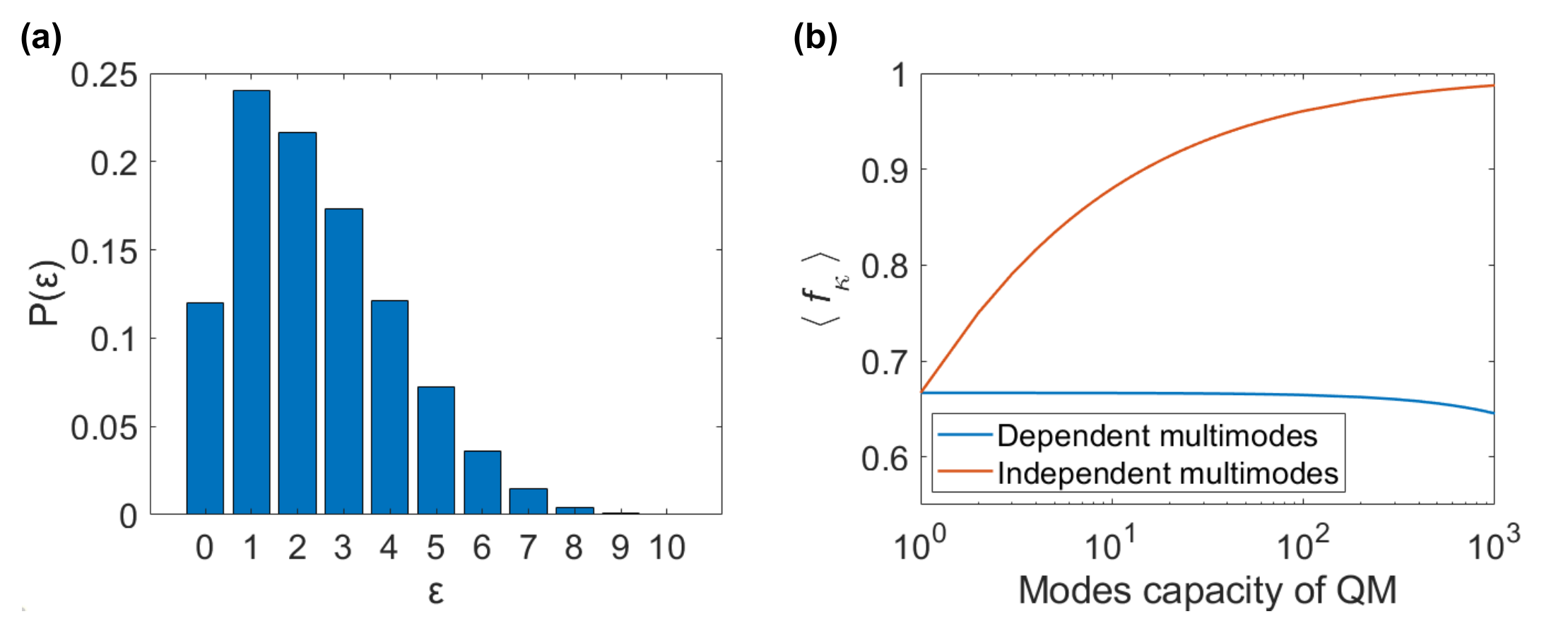}
\caption{Superiority of independent modes multiplexing.
\textbf{(a)} Probability distribution of residual entanglements number $\varepsilon$, with $n=10$ and $p=10^{-4}$.
\textbf{(b)} Normalized mode efficiency as a function of QM modes capacity with $p=10^{-4}$.
}
\label{fig:mn}
%label要放在最下面，否则hyperref包会报错！
\end{figure*}
 
Here we discuss the superiority of independent modes multiplexing. As discussed in the main text, we define the average mode efficiency in a two-segment system as:
\begin{equation}
    \langle f_\kappa \rangle=\frac{\langle T_{EG} \rangle}{\kappa\langle T \rangle P_1},
\label{eq.fk}
\end{equation}
where $\langle T_{EG} \rangle$ is the average time for successful entanglement generation in a single mode, and $\langle T \rangle P_1$ represents the average time for an attempt of entanglement swapping in $\kappa$ modes. This metric reflects the ratio of a QM mode's operational duration to the total time, reaching unity when both segments consistently herald successful entanglement generation in every system clock cycle. However, in practice, the success probability of entanglement generation is quite low, meaning that segment A, with prepared entanglement, must wait for a successful entanglement generation at segment B. The occupied QM mode in segment A remains idle until the next entanglement swapping, resulting in $\langle f_\kappa \rangle<1$.

The subscript $\kappa$ is a general representation for the number of modes,with $\kappa=m$ for dependent modes and $\kappa=n$ for independent modes. Normalized by system clock, the EDR becomes 
\begin{equation}
   f=\frac{1}{\langle T \rangle}=\kappa pP_1 \langle f_\kappa \rangle,
   \label{edr kappa f}
\end{equation}
where $p=\langle T_{EG} \rangle^{-1}$ represents the success probability of entanglement generation in a single mode. According to Eq.~\ref{edr kappa f}, as mode number scales, EDR increase as a function of $\kappa$ and $\langle f_\kappa \rangle$, where $\langle f_\kappa \rangle$ represents the nonlinear relationship between $f$ and $\kappa$.

Assuming a sufficiently high EPPS repetition rate and QM lifetime, for a two-segment quantum repeater with single-mode QMs, $\langle f_\kappa \rangle$ is derived in the Appendix A.2 in Ref.~\cite{RevModPhys.83.33}:
\begin{equation}
    \langle f_\kappa \rangle|_{\kappa=1}=\frac{2-p}{3-2p}, 
\label{eq:fk=1}
\end{equation}
where $\langle f_\kappa \rangle|_{\kappa=1}=1$ when $p=1$, as required. Conversely, when $p$ approaches 0, $\langle f_\kappa \rangle|_{\kappa=1}$ asymptotically approaches $\frac{2}{3}$, which leads to the factor of $\frac{3}{2}$ in the EDR of hierarchical quantum repeater schemes~\cite{simon_quantum_2007,PhysRevA.78.012350}.

For dependent multimodes, the multiplexing only improves the success probability of entanglement generation from $p$ to $p_m=1-(1-p)^m$. Thus, we have:
\begin{equation}
    \langle f_m \rangle=\frac{2-p_m}{3-2p_m}\frac{p_m}{mp}=\frac{1-(1-p)^{2m}}{mp\left[1+2(1-p)^m\right]}.
\end{equation}

For independent multimodes, we derive:
\begin{equation}
\langle f_n \rangle=\sum^{n}_{\varepsilon=0}\frac{2n-\varepsilon}{2n}P_n(\varepsilon),
\end{equation}
by substituting equation Eq.~\ref{Eq.T} into Eq.~\ref{eq.fk}.

% Figure~\ref{fig:mn} (a) illustrates the theoretical upper bound of EDR at that distance for different types of multiplexing, where independent modes multiplexing achieves about 1.5 times the EDR compared to dependent modes multiplexing with hundreds of modes.

In the specific condition of H4QR, $p$ is approximately $10^{-4}$ with a distribution distance of 3000 km.
Fig.~\ref{fig:mn} (a) shows a typical probability distribution $P_n(\varepsilon)$ with 10 independent multiplexing modes. The skewed $P_n(\varepsilon)$ distribution toward low $\varepsilon$ results in a higher $\langle f_n \rangle$ than $\langle f_n \rangle|_{n=1}$. 
Noting that $P_1(0)\approx\frac{1}{3}$ and $P_1(1)\approx\frac{2}{3}$, the mode efficiency $\langle f_n \rangle|_{n=1}\approx\frac{2}{3}$, consistent with the result of Eq.~\ref{eq:fk=1}. 

Fig.~\ref{fig:mn} (b) shows the $\langle f_\kappa \rangle$ as a function of $\kappa$ in different type of multiplexing. In the case of independent multimodes, $\langle f_n \rangle$ increases as $n$ increases, and approaches unity eventually. On the contrary, $\langle f_m \rangle$ is monotonically decreasing functioned with $m$, which leads to a greater gap between independent and dependent modes multiplexing as $\kappa$ grows. The underlying reason is that with an increasing number of independent modes $n$, the fraction of occupied modes becomes progressively smaller. Consequently, while dependent modes multiplexing can boost the entanglement generation probability $p_m$, it cannot increase the rate at which an occupied mode is released. Since $p_m$ has an upper limit of 1, the marginal gain from adding more dependent modes diminishes as the modes number grows large.

In hierarchical quantum repeater schemes, each entanglement swapping requires established entanglement on both sides, which leads to the total entanglement distribution time scales as $\langle f_\kappa \rangle^{-N}$ with nesting level $N$. This scaling causes the performance gap between independent and dependent multiplexing to grow substantially in multi-layer quantum repeaters.

%\section*{Author contributions}
%Z.-Q.Z. proposed the original protocol and supervised all aspects of this work; P.-X. L. improved the protocol and calculated the entanglement distribution time, Y.-P. L. improved the channel design and calculated the channel efficiency, P.-X. L., Y.-P. L. and Z.-Q.Z. wrote the manuscript with input from others. Z.-Q.Z. and C.-F.L. supervised the project. All authors discussed the calculation procedures and results.\\

%%%%%%%%%%%%%%%%%%%%%%%%%%%%%%%%%%%%%%%%%%%%%%%%%%%%%%%%
\bibliography{mainropp.bib}
\end{document}